\documentclass[12pt,a4paper]{article}
\usepackage[english]{babel}
\usepackage[utf8]{inputenc}

\usepackage{amsmath}
\usepackage{amssymb}
\usepackage{amsfonts}
\usepackage{amsbsy}

\usepackage{graphicx}

\usepackage{newlfont}
\usepackage{float}
\usepackage[paperwidth=192mm,
            paperheight=262mm,
            vmargin={19mm,19mm},
            hmargin={13.7mm,13.7mm},
            headsep=12pt,
            footskip=12pt]{geometry}

\usepackage{hyperref}

\usepackage{siunitx}
\usepackage{subcaption}

\usepackage{caption}
\usepackage{multirow}

\usepackage{booktabs}
\usepackage{tabularx}

\usepackage{placeins}

\usepackage[normalem]{ulem}

\usepackage[backend=bibtex,style=numeric,sorting=nyt]{biblatex}
\renewbibmacro{in:}{}
\DeclareMathOperator{\grad}{\nabla}
\DeclareMathOperator{\dive}{\nabla\cdot}

\begin{document}

\title{Robust and accurate simulations of flows over orography using non-conforming meshes}

\author{Giuseppe Orlando\(^{(1)}\), \\ 
        Tommaso Benacchio\(^{(2)}\), Luca Bonaventura\(^{(3)}\)}

\date{}

\maketitle

\begin{center}
{
\small
\(^{(1)}\)  
CMAP, CNRS, \'{E}cole polytechnique, Institut Polytechnique de Paris \\ Route de Saclay, 91120 Palaiseau, France \\
{\tt giuseppe.orlando@polytechnique.edu} \\
\ \\
\(^{(2)}\)  
Weather Research, Danish Meteorological Institute \\
Sankt Kjelds Plads 11, 2100 Copenhagen, Denmark \\
{\tt tbo@dmi.dk} \\
\ \\
\(^{(3)}\)  
Dipartimento di Matematica, Politecnico di Milano \\
Piazza Leonardo da Vinci 32, 20133 Milano, Italy \\
{\tt luca.bonaventura@polimi.it} \\
}
\end{center}

\noindent


\pagebreak

\abstract{We systematically validate the static local mesh refinement capabilities of a recently proposed IMEX-DG scheme implemented in the framework of the \texttt{deal.II} library. Non-conforming meshes are employed in atmospheric flow simulations to increase the resolution around complex orography. The proposed approach is fully mass and energy conservative and allows local mesh refinement in the vertical and horizontal direction without relaxation at the internal coarse/fine mesh boundaries. A number of numerical experiments based on classical benchmarks with idealized as well as more realistic orography profiles demonstrate that simulations with the locally refined mesh are stable for long lead times and that no spurious effects arise at the interfaces of mesh regions with different resolutions. Moreover, correct values of the momentum flux are retrieved and the correct large-scale orographic response is reproduced. Hence, large-scale orography-driven flow features can be simulated without loss of accuracy using a much lower total amount of degrees of freedom.}

\pagebreak

\section{Introduction}
\label{sec:intro} \indent

Atmospheric flows display phenomena on a very wide range of spatial scales that interact with each other. Many strongly localized features, such as complex orography or hurricanes, can only be modelled correctly if a very high spatial resolution is employed, especially in the lower troposphere, while larger scale features such as high/low pressure systems and stratospheric flows can be adequately resolved on much coarser meshes. The impact of orography on the atmospheric circulation has been the focus of a large number of studies, see e.g. the classical work \cite{mcfarlane:1987}, the more recent review \cite{sandu:2019}, and references therein. This impact is significant both on the short and on the long time scales and even affects the oceanic circulation \cite{maffre:2018}. The minimal resolution requirements for an accurate description of the atmospheric phenomena relevant for numerical weather prediction (NWP) and climate models have been subject to strong debate, see e.g. the classical paper \cite{lindzen:1989}, the more recent contribution \cite{skamarock:2019}, and the references therein. Furthermore, for numerical reasons, orography data used by NWP and climate models are often filtered, thus limiting the scales at which orography can effectively be represented in numerical models. For example, the analysis in \cite{davies:2001} showed that orographic features must be resolved by a sufficiently large number of mesh points (from 6 to 10) in finite difference models to avoid spurious numerical features. 

The insufficient resolution of orographic features is compensated in NWP and climate models by subgrid-scale orographic drag parameterizations \cite{miller:1989, palmer:1986}, which are essential for an accurate description of atmospheric flows with models using feasible resolutions, see again the discussion in \cite{sandu:2019}. The interplay between resolved and parameterized orographic effects is critical, since many operational models currently employ resolutions in the so-called `grey zone', for which some orographic effects are well resolved while others still require parameterization. Global simulations with the ECMWF-IFS NWP model without drag parameterization showed that the increase in forecast skill for increasing atmospheric resolution was chiefly due to the improved representation of the orography \cite{kanehama:2019}. When parameterizing the drag, the positive impact of the parameterization decreased as long as the model resolution increased. Finally, sharper orography representations also proved beneficial for simulations of mountain wave-driven middle atmosphere processes \cite{fritts:2022}.

Because of the multiscale nature of the underlying processes, NWP is an apparently ideal application area for adaptive numerical approaches. However, mesh adaptation strategies have only slowly found their way into the NWP literature, due to limitations of earlier numerical methods, concerns about the accuracy of variable resolution meshes for the correct representation of typical atmospheric wave phenomena, and the greater complexity of an efficient parallel implementation for non-uniform or adaptive meshes. The first approaches to variable local mesh refinement were based on the nesting concept, see e.g. \cite{harrison:1972, phillips:1973, zhang:1986}. Early attempts to introduce adaptive meshes in NWP were then presented in the seminal papers \cite{skamarock:1989, skamarock:1993}, while a review of earlier variable resolution approaches is presented in \cite{cote:1997}. The impact of variable resolution meshes on classical finite difference methods was analyzed in \cite{long:2011, vichnevetsky:1987}. More recently, methods allowing mesh deformation strategies were proposed in \cite{prusa:2003} and techniques to estimate the required resolution were presented in \cite{weller:2009}, while applications of block structured meshes were discussed in \cite{jablonowski:2009}. In all these papers, finite difference or finite volume methods were employed for the numerical approximation. High-order finite element methods have also been exploited as an ingredient of accurate adaptive methods. More specifically, hybrid continuous-discontinuous finite element techniques were employed in \cite{li:2021}. Discontinuous Galerkin (DG) finite element $h-$adaptive approaches were proposed in \cite{kopera:2014, muller:2013, yelash:2014}, while $p-$adaptive DG methods for NWP were introduced in \cite{tumolo:2015}. An $hp-$ adaptive DG method for mesoscale atmospheric modelling has recently been proposed in \cite{dolejsi:2024} and a fully unstructured 3D approach was presented in \cite{tissaoui:2023}. Finally, in \cite{dueben:2014}, the impact of mesh refinement on large scale geostrophic equilibrium and turbulent cascades influenced by the Earth's rotation was investigated. 

Operational or semi-operational NWP models exist that have local mesh refinement \cite{skamarock:2012} or nesting \cite{skamarock:2021} capabilities. Almost all the published results, however, either require some relaxation at the boundaries between coarse and fine regions \cite{mctaggart:2011, tang:2013} or perform vertical mesh refinement over the whole vertical span of the computational domain \cite{daniels:2016, mahalov:2009, mirocha:2017}. In \cite{hellsten:2021}, a full 3D nesting approach without boundary relaxation is presented, which is only tested on cases either without orography or without stratification, with additional restrictions on the lateral boundary conditions that can be applied in the case of purely vertical nesting.

In this work, we test a recently proposed adaptive IMEX-DG method \cite{orlando:2023b, orlando:2022, orlando:2023a} on a number of benchmarks for atmospheric flow over idealized and real orography. The proposed approach is fully mass and energy conservative and allows local refinement in the vertical and horizontal direction without the need to apply relaxation at the internal coarse/fine mesh boundaries. Through a quantitative assessment of non-conforming $h-$adaptation, we aim to show that simulations with adaptive meshes around orography can increase the accuracy of the local flow description without affecting the larger scales, thereby significantly reducing the overall number of degrees of freedom compared to uniform mesh simulations. The employed numerical approach combines accurate and flexible DG space discretization with an implicit-explicit (IMEX) time discretization, whose properties and theoretical justifications are discussed in detail in \cite{orlando:2023b, orlando:2022, orlando:2023a}. The adaptive discretization is implemented in the framework of the open-source numerical library \texttt{deal.II} \cite{arndt:2023, bangerth:2007}, which provides the non-conforming $h-$refinement capabilities exploited in the numerical simulation of flows over orography. The numerical results show that simulations with the refined meshes provide stable solutions with greater or comparable accuracy to those obtained with the uniform mesh. Importantly, no spurious reflections arise at internal boundaries separating mesh regions with different resolution and correct values for the vertical flux of horizontal momentum are retrieved. Both on idealised benchmarks and on test cases over realistic orographic profiles, simulations using non-conforming local mesh refinement correctly reproduce the larger scale, far-field orographic response, with meshes that are relatively coarse over most of the domain. This supports the idea that locally refined, non-conforming meshes can be an effective tool to reduce the dependence of NWP and climate models on parametrizations of orographic effects \cite{kanehama:2019, sandu:2019}.
 
The paper is structured as follows. The model equations and a short introduction to non-conforming meshes are presented in Section \ref{sec:modeleq}. The quantitative numerical assessment of non-conforming mesh refinement over orography in a number of idealized and real benchmarks is reported in Section \ref{sec:tests}. Some conclusions and considerations about open issues and future work are presented in Section \ref{sec:conclu}.

\section{The model equations}
\label{sec:modeleq} \indent

The fully compressible Euler equations of gas dynamics represent the most comprehensive mathematical model for atmosphere dynamics, see e.g. \cite{davies:2003, giraldo:2008, steppeler:2003}. Let \(\Omega \subset \mathbb{R}^{d}, 2 \le d \le 3\) be a simulation domain and denote by \(\mathbf{x}\) the spatial coordinates and by \(t\) the temporal coordinate. We consider the unsteady compressible Euler equations, written in conservation form as
\begin{eqnarray}\label{eq:euler_comp}
    \frac{\partial\rho}{\partial t} + \dive\left(\rho\mathbf{u}\right) &=& 0 \nonumber \\
    \frac{\partial\left(\rho\mathbf{u}\right)}{\partial t} + \dive\left(\rho\mathbf{u} \otimes \mathbf{u}\right) + \grad p &=& \rho\mathbf{g} \\
    \frac{\partial\left(\rho E\right)}{\partial t} + \dive\left[\left(\rho E + p\right)\mathbf{u}\right] &=& \rho \mathbf{g} \cdot \mathbf{u}, \nonumber
\end{eqnarray}
for \(\mathbf{x} \in \Omega\), \(t \in (0, T_{f}]\), supplied with suitable initial and boundary conditions. Here \(T_{f}\) is the final time, \(\rho\) is the density, \(\mathbf{u}\) is the fluid velocity, \(p\) is the pressure, and \(\otimes\) denotes the tensor product. Moreover, \(\mathbf{g} = -g\mathbf{k}\) represents the acceleration of gravity, with \(g = \SI{9.81}{\meter\per\second\squared}\) and \(\mathbf{k}\) denoting the upward pointing unit vector in the standard Cartesian frame of reference. The total energy \(\rho E\) can be rewritten as \(\rho E = \rho e + \rho k\), where \(e\) is the internal energy and \(k = \frac{1}{2}\|\mathbf{u}\|^{2}\) is the kinetic energy. We also introduce the specific enthalpy \(h = e + \frac{p}{\rho}\) and we notice that one can rewrite the energy flux as
\[\left(\rho E + p\right)\mathbf{u} = \left(e + k + \frac{p}{\rho}\right)\rho\mathbf{u} = \left(h + k\right)\rho\mathbf{u}.\]
Notice that the choice of the total energy density \(E\) as prognostic variable has been shown, at least empirically, to yield model formulations that do not require special well balancing techniques for flows under the action of gravity \cite{baldauf:2023}. The above equations are complemented by the equation of state (EOS) for ideal gases, given by \(p = \rho RT\), with \(R\) being the specific gas constant. For later reference, we define the Exner pressure \(\Pi\) as
\begin{equation}
    \Pi = \left(\frac{p_{0}}{p}\right)^{\frac{\gamma - 1}{\gamma}},
\end{equation}
with \(p_{0} = \SI[parse-numbers=false]{10^{5}}{\pascal}\) being a reference a pressure and \(\gamma\) denoting the isentropic exponent. We consider \(\gamma = 1.4\) and the gas constant \(R = \SI{287}{\joule\per\kilogram\per\kelvin}\) for all the test cases.

\subsection{Non-conforming meshes}
\label{ssec:non_conforming_meshes}

We solve system \eqref{eq:euler_comp} numerically using the IMEX-DG solver proposed in \cite{orlando:2023b, orlando:2022} and validated in \cite{orlando:2023a} for atmospheric applications (see also \cite{orlando:2024}). While on one the hand no special well balancing property with respect to hydrostatic equilibrium has been proven for the proposed discretization (see e.g. the proposal in \cite{blaise:2016}), no evidence of numerical problems related to the representation of hydrostatic equilibrium was found in the many numerical tests performed in the previously mentioned papers. This could be related to the choice of the energy density as prognostic variable, as argued in \cite{baldauf:2023}, based on numerical results obtained with a similar formulation.

Atmospheric flows such as those considered in this work are characterized by low Mach numbers, as motions of interest have characteristic speeds much lower than that of sound. In the low Mach limit, terms related to pressure gradients yield stiff components in the system of ordinary differential equations resulting from the spatial discretization of system \eqref{eq:euler_comp}. Therefore, an implicit coupling between the momentum balance and the energy balance is adequate, whereas the density can be treated in a fully explicit fashion, see e.g. the discussion in \cite{casulli:1984, dumbser:2016}. The time discretization is based on a variant of the IMEX method proposed in \cite{giraldo:2013}, while the space discretization adopts the DG scheme implemented in the \texttt{deal.II} library \cite{arndt:2023}. We refer to \cite{orlando:2023b, orlando:2022, orlando:2023a} for a complete analysis and discussion of the numerical methodology, and to \cite{giraldo:2020} for a comprehensive introduction to the DG method.

The nodal DG method, as the one employed in \texttt{deal.II} \cite{arndt:2023}, is characterized by integrals over faces belonging to two elements. Moreover, a weak imposition of boundary conditions is typically adopted \cite{arnold:2002}. Hence, the method provides a natural framework for formulations on multi-block meshes. Consider a generic non-linear conservation law
\begin{equation}
    \frac{\partial\Psi}{\partial t} + \dive\mathbf{F}(\Psi) = 0.
\end{equation}
We multiply the previous relation by a test function \(\Lambda\) and, after integration by parts, we obtain the following local formulation on each element \(K\) of the mesh with boundary \(\partial K\):
\begin{equation}\label{eq:weak_form}
    \int_{K} \frac{\partial\Psi}{\partial t}\Lambda d\Omega - \int_{K} \mathbf{F}(\Psi) \cdot \nabla\Lambda d\Omega + \int_{\partial K} \hat{\mathbf{F}}(\Psi^{+}, \Psi^{-})\Lambda d\Sigma = 0,
\end{equation}
where \(d\Omega\) is the volume element and \(d\Sigma\) is the surface element. In the surface integral, we replace the term \(\mathbf{F}(\Psi)\) with a numerical flux \(\hat{\mathbf{F}}(\Psi^{+}, \Psi^{-})\), which depends on the solution on both sides \(\Psi^{+}\) and \(\Psi^{-}\) of an interior face. 

\begin{figure}[h!]
    \centering
    \includegraphics[width=0.5\textwidth]{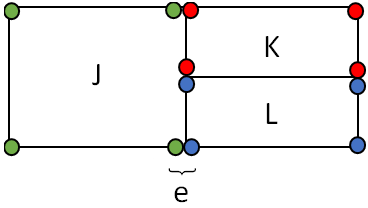}
    \caption{Example of two neighbouring cells in a non-conforming mesh. The two green nodes from cell \(J\), the two red nodes from cell \(K\) and the two blue nodes from cell \(L\) are involved in the computation of the flux in the boundary integral of \eqref{eq:weak_form} for face \(e\).}
    \label{fig:non_conforming_mesh}
\end{figure}

A non-conforming mesh is characterized by cells with different refinement levels, so that the resolution between two neighbouring cells can be different (Figure \ref{fig:non_conforming_mesh}). For faces between cells of different refinement level, the integration is performed from the refined side and a suitable interpolation is performed on the coarse side, so as to guarantee the conservation property, see the discussion in \cite{bangerth:2007}. Hence, no hanging nodes appear in the implementation of the discrete weak form of the equations. 

DG methods with non-conforming meshes have been developed for different applications, see e.g. \cite{fahs:2015, heinz:2023}. The main constraint posed by the \texttt{deal.II} library for the use of non-conforming meshes is the requirement of not having neighbouring cells with refinement levels differing by more than one. Thus, for each non-conforming face, flux contributions have to be considered at most from two refined faces in two dimensions and from four faces in three dimensions (Figure \ref{fig:non_conforming_mesh}). The out-of-the-box availability in \texttt{deal.II} provides an ideal testbed for evaluating the potential computational savings using non-conforming meshes instead of uniform meshes in atmospheric flow simulations. 

\section{Numerical results}
\label{sec:tests} \indent
 
We consider a number of benchmarks of atmospheric flows over orography for the validation of NWP codes, see e.g. the seminal papers \cite{klemp:1983, klemp:1978} and results and discussions in \cite{bonaventura:2000, melvin:2019, pinty:1995, tumolo:2015}. The objective of these tests is twofold. First, we evaluate the stability and accuracy of numerical solutions obtained using non-conforming meshes compared to those obtained using uniform meshes. Second, we assess the computational cost carried by both setups and potential advantages at a given accuracy level.

Discrete parameter choices for the numerical simulations are associated with two Courant numbers, the acoustic Courant number \(C\), based on the speed of sound \(c\), and the advective Courant number \(C_{u}\), based on the magnitude of the local flow velocity \(u\): 
\begin{equation}\label{eq:Courant}
    C = rc\Delta t/{\mathcal{H}}  \qquad  C_{u} = ru\Delta t/{\mathcal{H}}.
\end{equation}
Here \(r\) is the polynomial degree used for the DG spatial discretization, \(\mathcal{H}\) is the minimum cell diameter of the computational mesh, and \(\Delta t\) is the time-step adopted for the time discretization. We consider polynomial degree \(r = 4\), unless differently stated. Wall boundary conditions are employed for the bottom boundary, whereas non-reflecting boundary conditions are required by the top boundary and the lateral boundaries. For this purpose, we introduce the following Rayleigh damping profile \cite{melvin:2019, orlando:2023a}:
\begin{equation}
    \lambda = 
    \begin{cases}
        0, \qquad & \text{if } z < z_{B} \\
        \overline{\lambda}\sin^2\left[\frac{\pi}{2}\left(\frac{z - z_{B}}{z - z_{T}}\right)\right] \qquad & \text{if } z \ge z_{B},
    \end{cases} 
\end{equation}
where \(z_{B}\) denotes the height at which the damping starts and \(z_{T}\) is the top height of the considered domain. Analogous definitions apply for the two lateral boundaries. The classical Gal-Chen height-based terrain-following coordinate \cite{galchen:1975} is used to obtain a terrain-following mesh in Cartesian coordinates.

A relevant diagnostic quantity to check that a correct orographic response is achieved is represented by the vertical flux of horizontal momentum (henceforth ``momentum flux''), defined as \cite{smith:1979}
\begin{equation}\label{eq:momentum_flux_def}
    m(z) = \int_{-\infty}^{\infty} \overline{\rho}(z)u'(x,z)w'(x,z)dx.
\end{equation} 
Here, \(u'\) and \(w'\) denote the deviation from the background state of the horizontal and vertical velocity, respectively. Table \ref{tab:parameters} reports the parameters employed for the different test cases.

\begin{table}[h!]
    \centering
    \footnotesize
    \begin{tabularx}{0.95\columnwidth}{lrrrXXrrr}
        \toprule
        \textbf{Test case} & \(\boldsymbol{\Delta}\mathbf{t}\) & \(\mathbf{T_{f}}\) & \textbf{Domain} & \textbf{Damping} & \textbf{Damping} & \(\overline{\boldsymbol{\lambda}}\boldsymbol{\Delta}\mathbf{t}\) & \(\mathbf{C}\) & \(\mathbf{C_{u}}\) \\
        & \(\boldsymbol{[}\SI{}{\second}\boldsymbol{]}\) & \(\boldsymbol{[}\SI{}{\hour}\boldsymbol{]}\) & \(\boldsymbol{[}\SI{}{\kilo\meter} \times \SI{}{\kilo\meter}\)\(\boldsymbol{]}\) & \textbf{layer} \(\boldsymbol{(}x\boldsymbol{)}\) & \textbf{layer} \(\boldsymbol{(}z\boldsymbol{)}\) & & &  \\
        \midrule
        LHMW & 2.5 & 15 & \(240 \times 30\) & (0,80), (160,240) & (15,30) & 0.3 & 3.66 & 0.23 \\
        \midrule
        NLNHMW & 1 & 5 & \(40 \times 20\) & (0,10), (30,40) & (9,20) & 0.15 & 2.16 & 0.13 \\ 
        \midrule
        BWS & 0.75 & 3 & \(220 \times 25\) & (0,30), (190,220) & (20,25) & 0.15 & 0.79 & 0.23 \\
        \midrule
        T-REX & 0.75 & 4 & \(400 \times 26\) & (0,50), (350,400) & (20,26) & 0.15 & 1.34 & 0.29 \\
        \bottomrule
    \end{tabularx}
    \caption{Model parameters for the two-dimensional test cases in this Section, see main text for details. LHMW: linear hydrostatic mountain wave. NLNHMW: nonlinear nonhydrostatic mountain wave. BWS: Boulder windstorm (inviscid configuration). T-REX: Sierra profile, T-REX experiment. The intervals where the damping layers are applied are in units of km.}
    \label{tab:parameters}
\end{table}

\subsection{Linear hydrostatic flow over a hill}
\label{ssec:hydrostatic} \indent
 
\begin{figure}[h!]
    \centering
    \includegraphics[width=0.7\textwidth]{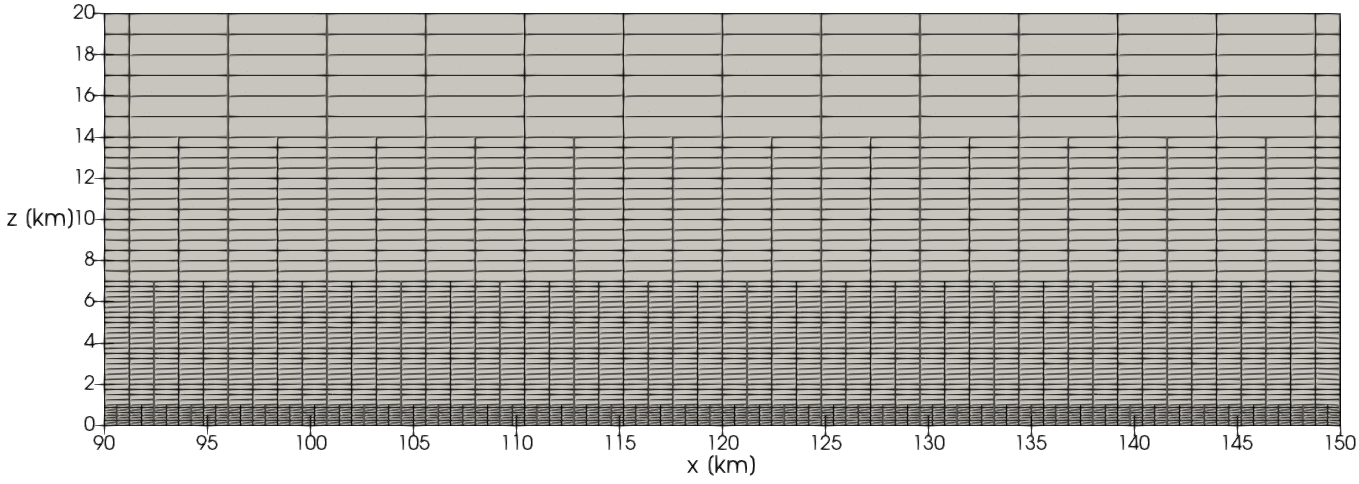}
    \caption{Linear hydrostatic flow over a hill, non-conforming mesh.}
    \label{fig:linear_hydro_non_conforming_mesh}
\end{figure}

\begin{figure}[h!]
    \centering
    \includegraphics[width=0.7\textwidth]{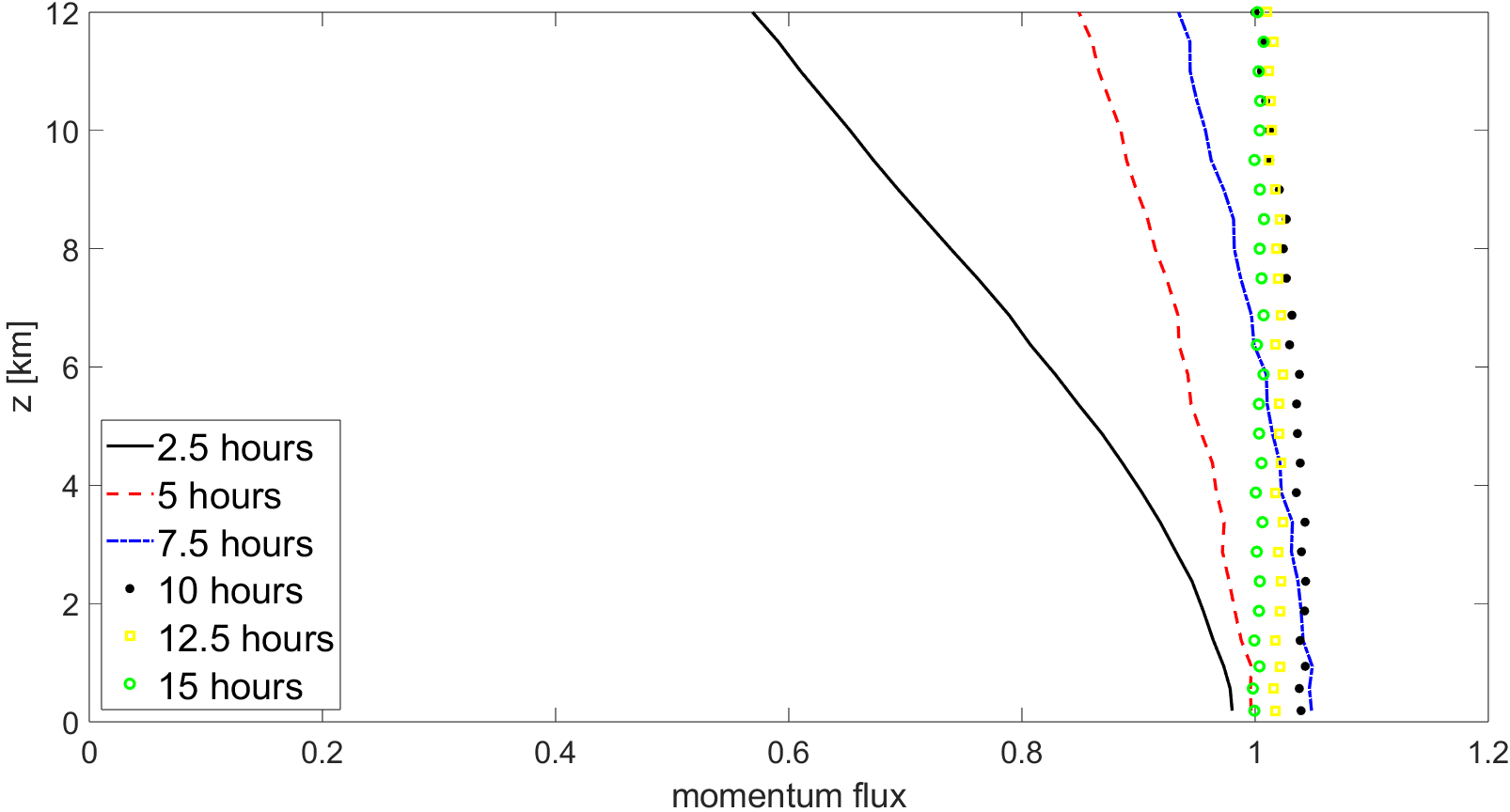}
    \caption{Linear hydrostatic flow over a hill, evolution of normalized momentum flux for the simulation using a non-conforming mesh.}
    \label{fig:linear_hydro_momentum}
\end{figure}

First, we consider a linear hydrostatic configuration, see e.g. \cite{giraldo:2008, orlando:2023a}. The bottom boundary is described by the function
\begin{equation}\label{eq:versiera_Agnesi}
    h(x) = \frac{h_{c}}{1 + \left(\frac{x - x_{c}}{a_{c}}\right)^2},
\end{equation}
the so-called \textit{versiera of Agnesi}, with \(h_{c}\) being the height of the hill and \(a_{c}\) being its half-width. We take \(h_{c} = \SI{1}{\meter}, x_{c} = \SI{120}{\kilo\meter}\), and \(a_{c} = \SI{10}{\kilo\meter}\). The initial state of the atmosphere consists of a constant horizontal flow with \(\overline{u} = \SI{20}{\meter\per\second}\) and of an isothermal background profile with temperature \(\overline{T} = \SI{250}{\kelvin}\) and Exner pressure:
\begin{equation}
    \overline{\Pi} = \exp\left(-\frac{g}{c_{p}\overline{T}}z\right),
\end{equation}
with \(c_{p} = R\frac{\gamma}{\gamma - 1}\) denoting the specific heat at constant pressure. In an isothermal configuration the Brunt-V{\"{a}}is{\"{a}}l{\"{a}} frequency is given by \(N = \frac{g}{\sqrt{c_{p}\overline{T}}}\). Hence, one can easily verify that
\begin{equation}
    \frac{N a_{c}}{\overline{u}} \gg 1,
\end{equation}
meaning that we are in a hydrostatic regime \cite{giraldo:2008, pinty:1995}. The computational mesh is composed by \(N_{el} = 1116\) elements with 4 different refinement levels (Figure \ref{fig:linear_hydro_non_conforming_mesh}). The finest level corresponds to a resolution of \(\SI{300}{\meter}\) along \(x\) and of \(\SI{62.5}{\meter}\) along \(z\), whereas the coarsest level corresponds to a resolution of \(\SI{2400}{\meter}\) along \(x\) and of \(\SI{500}{\meter}\) along \(z\). From the linear theory, the analytical momentum flux is given by \cite{smith:1979}
\begin{equation}\label{eq:analytic_momentum_hydro}
    m^{H} = -\frac{\pi}{4}\overline{\rho}_{s}\overline{u}_{s} N h_{c}^{2},
\end{equation}
with \(\overline{\rho}_{s}\) and \(\overline{u}_{s}\) denoting the surface background density and velocity, respectively. The computed momentum flux normalized by \(m^{H}\) approaches \(1\) as the simulation reaches the steady state (Figure \ref{fig:linear_hydro_momentum}). The momentum is correctly transferred in the vertical direction and no spurious oscillations arise at the interface between different mesh levels.

\begin{figure}[h!]
    \begin{subfigure}{\textwidth}
	\centering
        \includegraphics[width=0.7\textwidth]{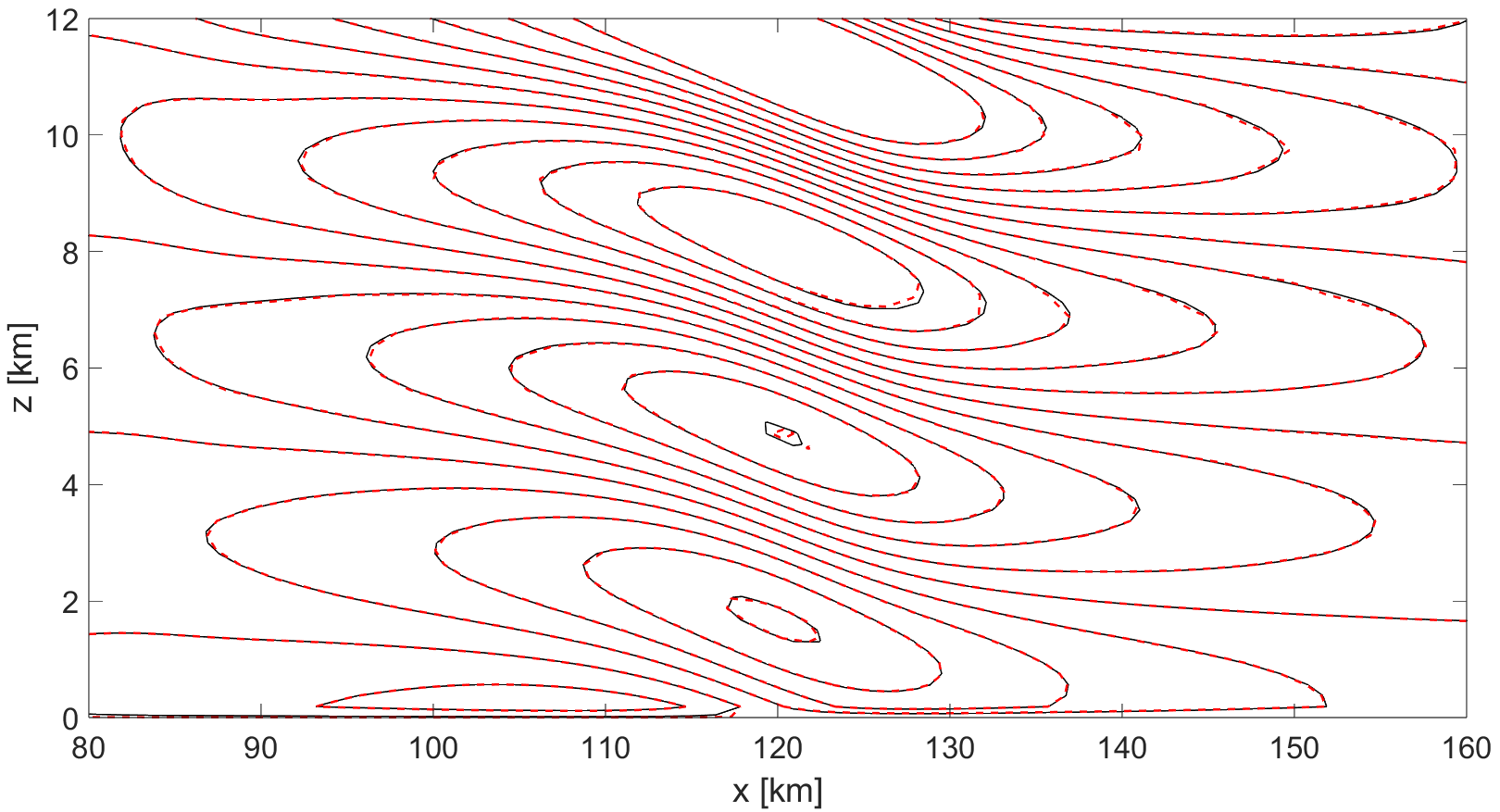}
    \end{subfigure}
    \begin{subfigure}{\textwidth}
	\centering
        \includegraphics[width=0.7\textwidth]{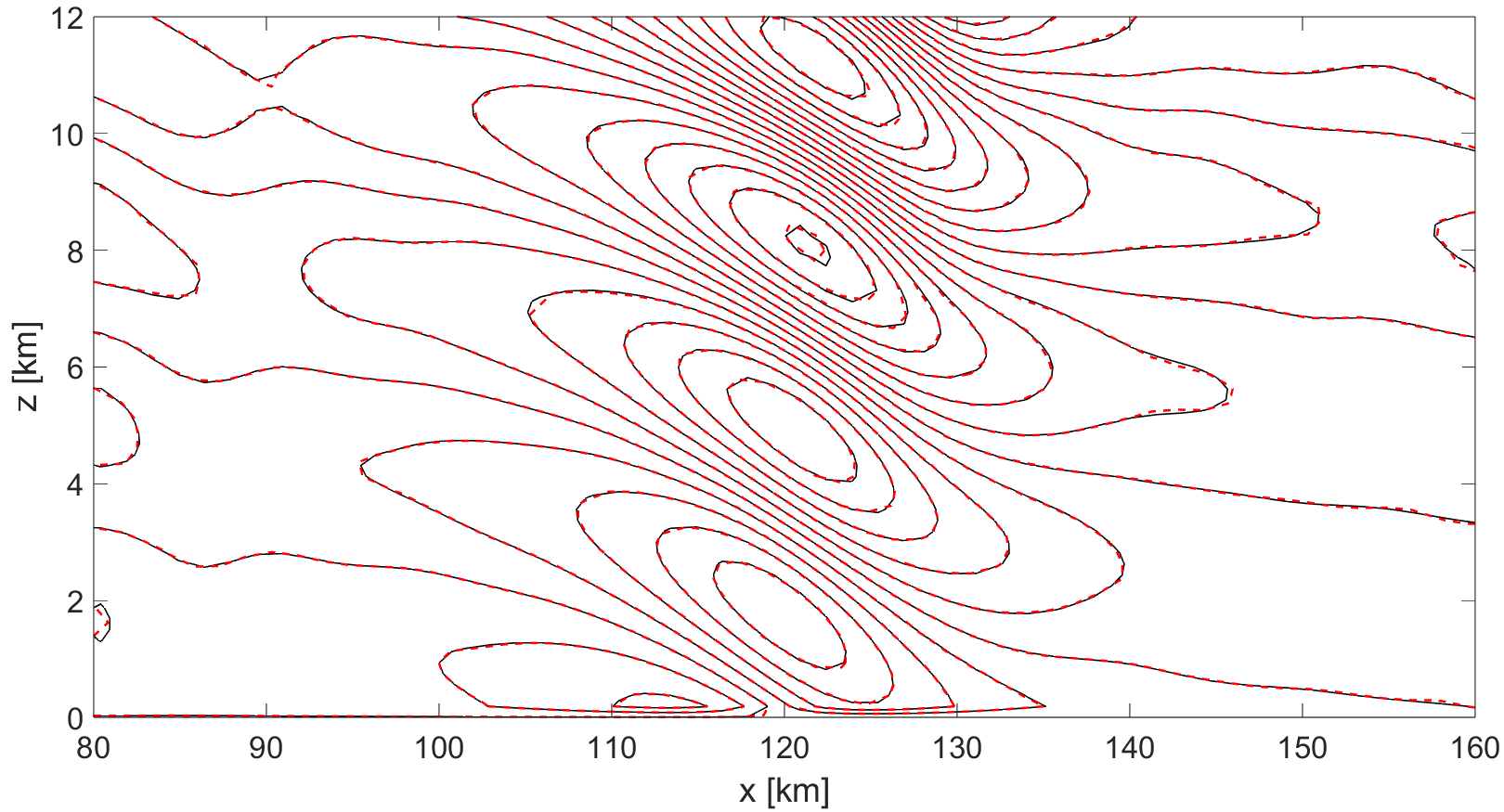}
    \end{subfigure}
    \caption{Linear hydrostatic flow over a hill at \(t = T_{f} = \SI{15}{\hour}\), numerical solutions using a non-conforming mesh (solid black lines) and the finest uniform mesh (dashed red lines). Top: horizontal velocity deviation, contours in \(\SI[parse-numbers=false]{[-2.5,2.5] \cdot 10^{-2}}{\meter\per\second}\) with a \(\SI[parse-numbers=false]{5 \cdot 10^{-3}}{\meter\per\second}\) contour interval; Bottom: vertical velocity, contours in \(\SI[parse-numbers=false]{[-4,4] \cdot 10^{-2}}{\meter\per\second}\) with a \(\SI[parse-numbers=false]{5 \cdot 10^{-4}}{\meter\per\second}\) contour interval.}
    \label{fig:linear_hydro_contours_comparison}
\end{figure}

A reference solution is computed using a uniform mesh with the maximum resolution of the non-conforming mesh, namely a mesh composed by \(200\) elements along the horizontal direction and \(120\) elements along the vertical one. A comparison of contour plots for the horizontal velocity deviation and for the vertical velocity shows an excellent agreement in the lee waves simulation between the finest uniform mesh and the non-conforming mesh (Figure \ref{fig:linear_hydro_contours_comparison}).

\begin{figure}[h!]
    \centering
    \includegraphics[width=0.7\textwidth]{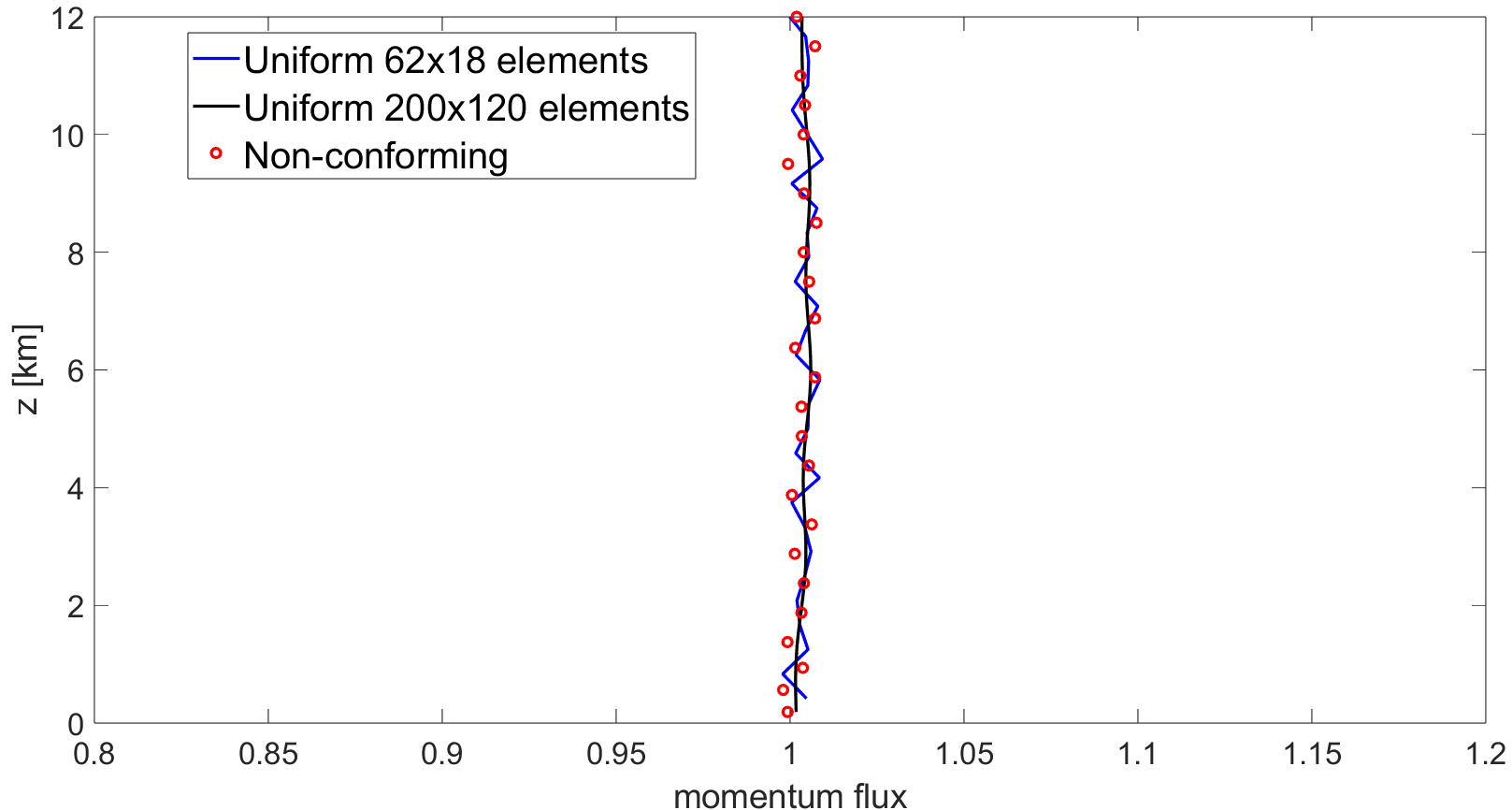}
    \caption{Linear hydrostatic flow over a hill, comparison of normalized momentum flux at \(t = T_{f} = \SI{15}{\hour}\) obtained using a uniform mesh at fine resolution (solid black line), a uniform mesh with the same number of elements as the non-conforming mesh (solid blue line), and the non-conforming mesh (red dots). See text for mesh definitions.}
    \label{fig:linear_hydro_momentum_comparison}
\end{figure}

In order to further emphasize the results obtained with the use of the non-conforming mesh, we consider a uniform mesh with the same number of elements (\(62 \times 18 = 1116\)) of the non-conforming mesh. A three-way comparison of the normalized momentum computed with the fine uniform mesh, the coarse uniform mesh, and the non-conforming mesh shows a good agreement with the analytical solution \(m^{H}\) \ref{fig:linear_hydro_momentum_comparison}). From a more quantitative point of view, we compute relative errors with respect to the reference solution in the portion of the domain \(\Omega = \SI[parse-numbers=false]{\left[80, 160\right]}{\kilo\meter} \times \SI[parse-numbers=false]{\left[0, 12\right]}{\kilo\meter}\) (Table \ref{tab:relative_errors_linear_hydro}). Moreover, we consider different configurations with non-conforming meshes and we compare them with configurations employing a uniform mesh using the same number of elements. Non-conforming mesh simulations significantly outperform uniform mesh simulations in terms of accuracy at a given number of degrees of freedom. At the finest \(\SI{300}{\meter}\) horizontal resolution and \(\SI{62.5}{\meter}\) vertical resolution, the use of the non-conforming mesh leads to a computational time saving of around \(90\%\) over the corresponding uniform mesh (bold numbers in Table \ref{tab:relative_errors_linear_hydro}). However, the present non-conforming mesh implementation is instead less competitive considering the wall-clock time at a given number of elements. This is due to the fact that, on non-conforming meshes, the condition number of the linear systems resulting from the IMEX discretization increases substantially, leading to a higher number of iterations for the GMRES solver \cite{du:2009, kamenski:2014, orlando:2022}. While some effective geometric multigrid preconditioners are available for non-symmetric systems arising from elliptic equations \cite{bramble:1994, esmaily:2018}, their extension to hyperbolic problems and their implementation in the context of the matrix-free approach of the \texttt{deal.II} library is not straightforward and will be the subject of future work. 

\begin{table}[h!]
    \centering
    \footnotesize
    \begin{tabularx}{0.95\columnwidth}{rXXXrcXXXrr}
        \toprule
        \multirow{2}{*}{\(N_{el}\)} & \multicolumn{3}{c}{Uniform} & & & & \multicolumn{3}{c}{Non-conforming} & \\
        \cmidrule(l){2-5}\cmidrule(l){7-11}
        & \(\Delta x[\SI{}{\meter}]\) & \(\Delta z[\SI{}{\meter}]\) & \(m(z)\) error & WT\([\SI{}{\second}]\) & & \(\Delta x_{\textrm{min}}[\SI{}{\meter}]\) & \(\Delta z_\textrm{min}[\SI{}{\meter}]\) & \(m(z)\) error & WT\([\SI{}{\second}]\) & Speed-up \\
        \midrule
        402 & 895.52 & 1250 & \num{6.10e-2} & 3050 & & 1200 & 250 &\num{4.34e-3} & 4210 & \\
        \midrule
        504 & 952.38 & 937.5 & \num{1.96e-2} & 3180 & & 600 & 125 & \num{2.34e-3} & 6510 & \\
        \midrule
        1116 & 967.74 & 416.67 & \num{3.94e-3} & 4470 & & 300 & 62.5 & \num{2.53e-3} & \textbf{14800} & \textbf{8.9} \\
        \midrule
        24000 & 300 & 62.5 & \multicolumn{1}{c}{-} & \textbf{131600} & & \multicolumn{1}{c}{-} & \multicolumn{1}{c}{-} & \multicolumn{1}{c}{-} & \multicolumn{1}{c}{-} & \\
        \bottomrule
    \end{tabularx}
    \caption{Linear hydrostatic flow over a hill: horizontal resolution \(\Delta x\), vertical resolution \(\Delta z\), \(l^{2}\) relative errors on the momentum flux, and wall-clock times (WT) for the uniform meshes and the non-conforming meshes. Here, and in the following tables, \(N_{el}\) denotes the number of elements. The speed-up is computed considering the same maximum spatial resolution, i.e. comparing the wall-clock time of the finest uniform mesh and the wall-clock time of the coarsest non-conforming mesh (bold WT, see also main text for further details).}
    \label{tab:relative_errors_linear_hydro}
\end{table}

\subsection{Nonlinear non-hydrostatic flow over a hill}
\label{ssec:nonhydrostatic} \indent

\begin{figure}[h!]
    \centering
    \includegraphics[width=0.7\textwidth]{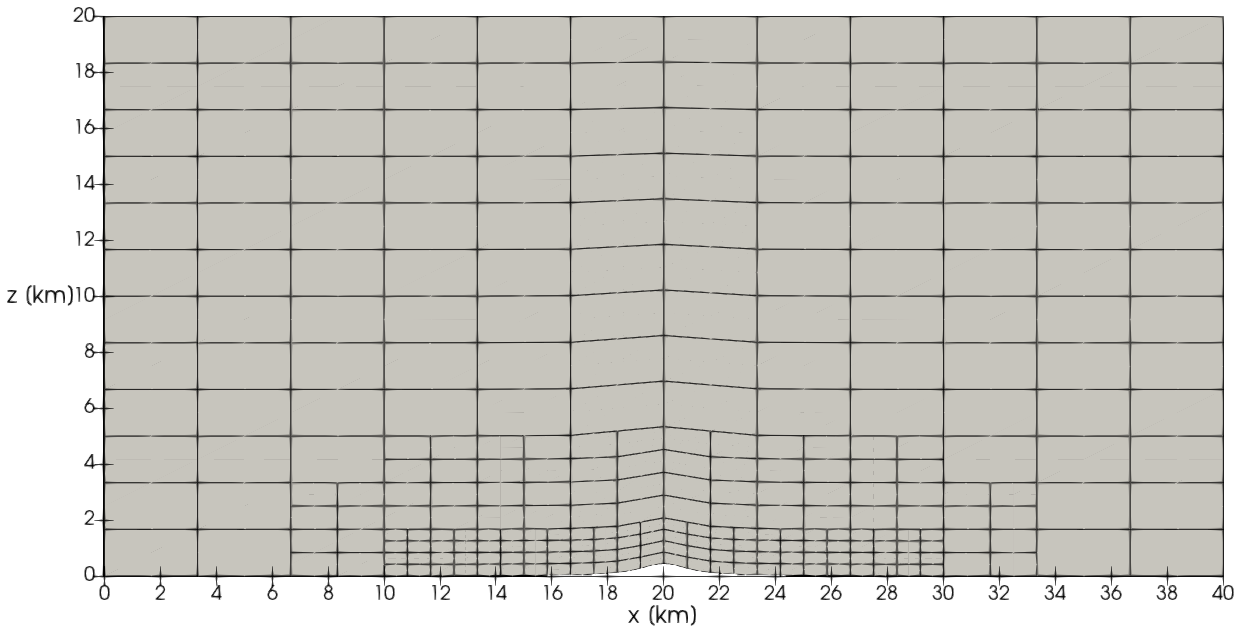}
    \caption{Nonlinear non-hydrostatic flow over a hill, non-conforming mesh.}
    \label{fig:nonlinear_nonhydro_non_conforming_mesh}
\end{figure}

Next, we consider a non-hydrostatic regime for which 
\begin{equation}
    \frac{N a_{c}}{\overline{u}} \approx 1. 
\end{equation}
More specifically, we focus on a nonlinear non-hydrostatic case \cite{orlando:2023a, restelli:2007, tumolo:2015}. The bottom boundary is described again by \eqref{eq:versiera_Agnesi}, with \(h_{c} = \SI{450}{\meter}, x_{c} = \SI{20}{\kilo\meter}\), and \(a_{c} = \SI{1}{\kilo\meter}\). The initial state of the atmosphere is described by a constant horizontal flow with \(\overline{u} = \SI{13.28}{\meter\per\second}\) and by the following potential temperature and Exner pressure:
\begin{eqnarray}
    \overline{\theta} &=& \theta_{ref}\exp\left(\frac{N^2}{g}z\right) \\
    \overline{\Pi} &=& 1 + \frac{g^{2}}{c_{p}\theta_{ref}N^2}\left[\exp\left(-\frac{N^{2}}{g}z\right) - 1\right],
\end{eqnarray}
with \(\theta_{ref} = \SI{273}{\kelvin}\) and \(N = \SI{0.02}{\per\second}\). The mesh is composed by \(N_{el} = 282\) elements with \(3\) different resolution levels (Figure \ref{fig:nonlinear_nonhydro_non_conforming_mesh}). The finest level corresponds to a resolution of \(\SI{208.33}{\meter}\) along \(x\) and of \(\SI{104.17}{\meter}\) along \(z\), whereas the coarsest level corresponds to a resolution of around \(\SI{833.33}{\meter}\) along \(x\) and of \(\SI{416.67}{\meter}\) along \(z\).

\begin{figure}[h!]
    \begin{subfigure}{\textwidth}
	\centering
        \includegraphics[width=0.7\textwidth]{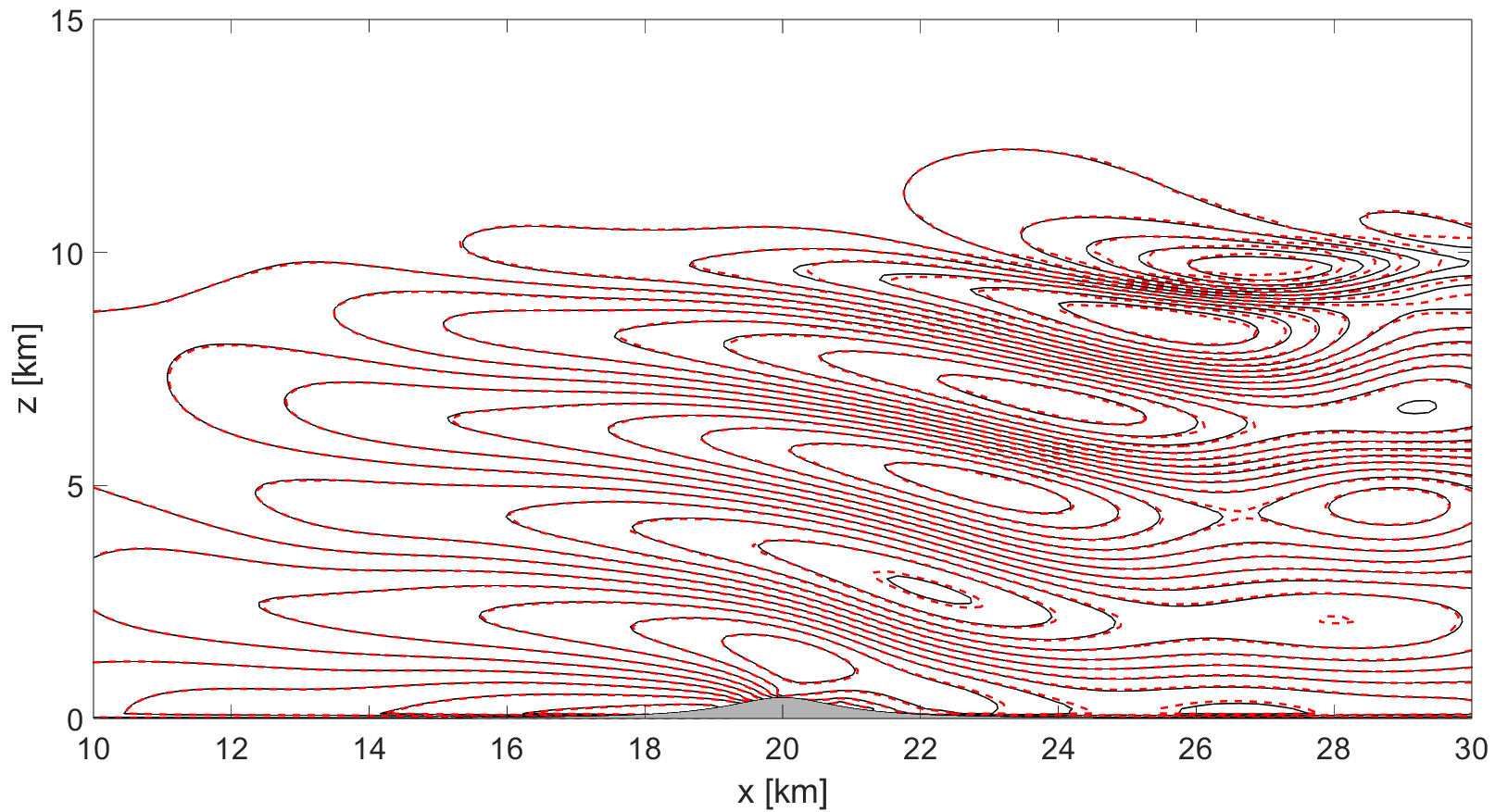}
    \end{subfigure}
    \begin{subfigure}{\textwidth}
	\centering
        \includegraphics[width=0.7\textwidth]{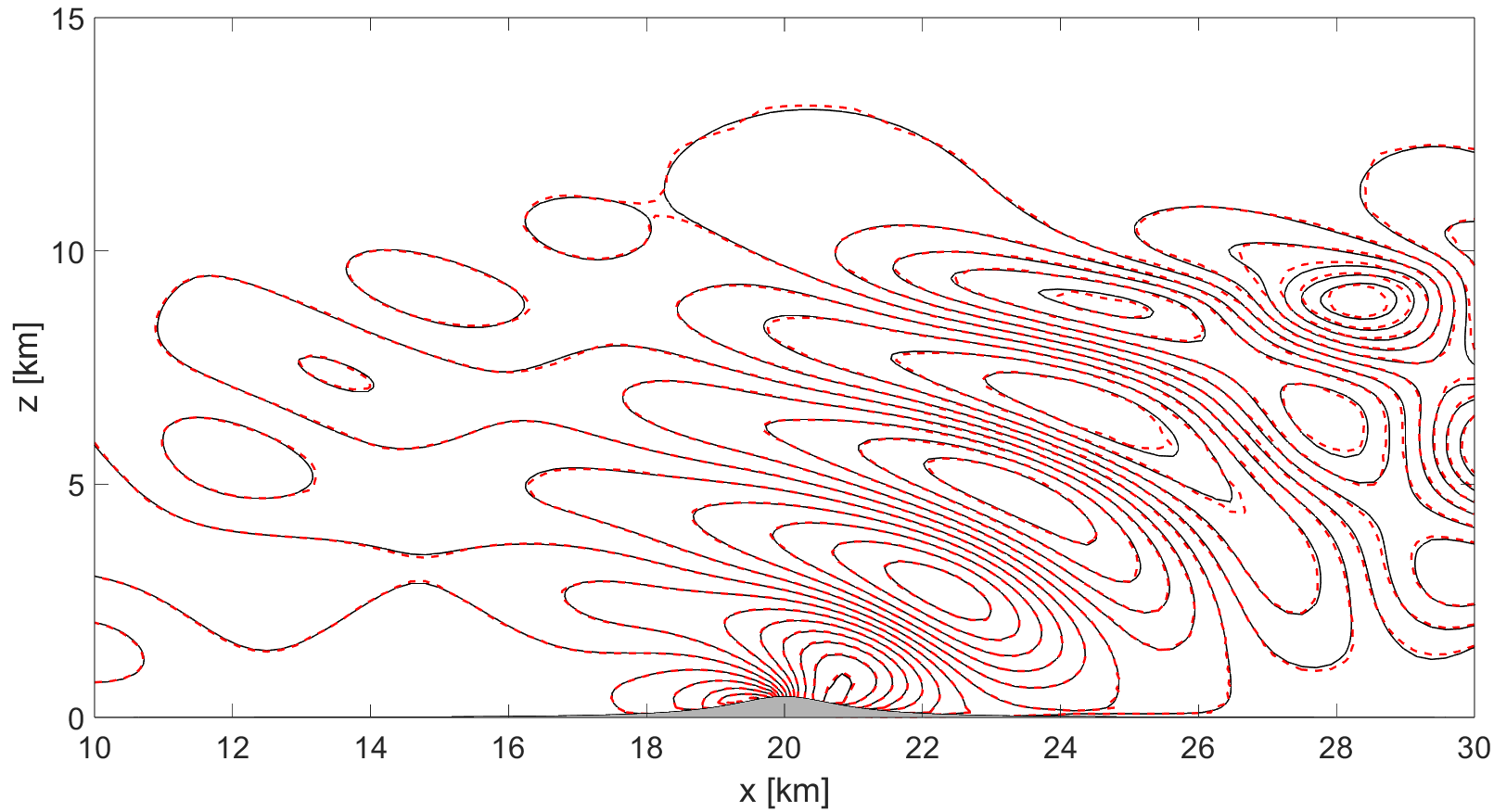}
    \end{subfigure}	
    \caption{Nonlinear non-hydrostatic flow over a hill at \(t = T_{f} = \SI{5}{\hour}\), computed on a uniform mesh at finer resolution (solid black lines) and on a non-conforming mesh (dashed red lines). Top: horizontal velocity deviation, contours in the interval \(\SI[parse-numbers=false]{[-7.2, 9.0]}{\meter\per\second}\) with a \(\SI{1.16}{\meter\per\second}\) interval. Bottom: vertical velocity, contours in the interval \(\SI[parse-numbers=false]{[-4.2, 4.0]}{\meter\per\second}\) with a \(\SI{0.586}{\meter\per\second}\) interval.}
    \label{fig:nonlinear_nonhydro_contours_comparison}
\end{figure}

\begin{figure}[h!]
    \centering
    \includegraphics[width=0.7\textwidth]{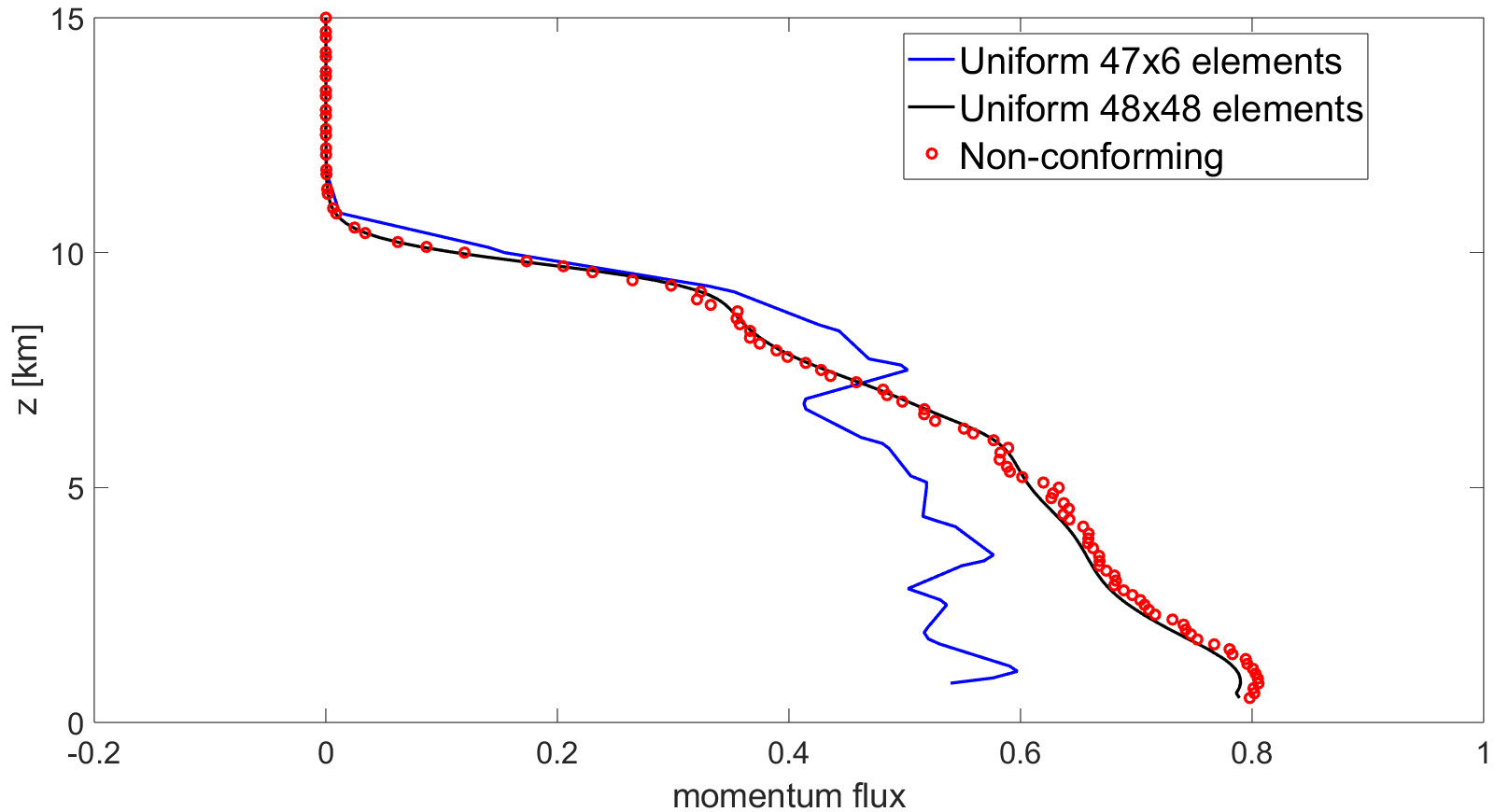}
    \caption{Nonlinear non-hydrostatic flow over a hill, comparison of normalized momentum flux at \(t = T_{f} = \SI{5}{\hour}\) obtained using a uniform mesh at fine resolution (black solid line), a uniform mesh with the same number of elements of the non-conforming mesh (blue solid line), and the non-conforming mesh (red dots).}
    \label{fig:nonlinear_nonhydro_momentum_comparison}
\end{figure}

A reference solution is computed using a uniform mesh with \(48 \times 48 = 2304\) elements, which corresponds to the finest resolution of the non-conforming mesh. A comparison of contour plots for the horizontal velocity deviation and for the vertical velocity shows good agreement between the finest uniform mesh and the non-conforming mesh in the development of lee waves (Figure \ref{fig:nonlinear_nonhydro_contours_comparison}). The use of the non-conforming mesh yields a computational time saving of around \(60\%\) (bold numbers in Table \ref{tab:relative_errors_nonlinear_nonhydro}). In addition, we consider a mesh with uniform resolution and with the same number of elements \(47 \times 6 = 282\) of the non-conforming mesh. We compare the computed normalized momentum flux at \(t = T_{f}\) in the reference configuration, in the non-conforming mesh configuration, and in the configuration with a uniform mesh with the same number of elements of the non-conforming mesh (Figure \ref{fig:nonlinear_nonhydro_momentum_comparison}). In terms of relative error with respect to the reference solution, the locally refined non-conforming mesh outperforms the uniform mesh by about an order of magnitude using the same number of elements (Table \ref{tab:relative_errors_nonlinear_nonhydro}). Analogous considerations to those reported in Section \ref{ssec:hydrostatic} are valid for the computational time.

\begin{table}[h!]
    \centering
    \footnotesize
    \begin{tabularx}{0.9\columnwidth}{rXXXrr}
	\toprule
        \(N_{el}\) & \(\Delta x[\SI{}{\meter}]\) & \(\Delta z[\SI{}{\meter}]\) & \(m(z)\) error & WT\([\SI{}{\second}]\) & Speed-up \\
	\midrule
	282 (uniform) & 212.77 & 833.33 & \num{2.31e-1} & 4520 & \\
	\midrule
        282 (non-conforming) & 208.33 & 104.17 & \num{1.92e-2} & \textbf{9680} & \textbf{2.3} \\
	\midrule
        2304 (uniform) & 208.33 & 104.17 & \multicolumn{1}{c}{-} & \textbf{22500} & \\
	\bottomrule
    \end{tabularx}
    \caption{Nonlinear non-hydrostatic flow over a hill: horizontal resolution \(\Delta x\), vertical resolution \(\Delta z\), \(l^{2}\) relative errors on the momentum flux, and wall-clock times (WT) for the uniform meshes and the non-conforming mesh. The speed-up is computed comparing the wall-clock time of the finest uniform mesh and the wall-clock time of the non-conforming mesh, which have the same resolution (see also main text for further details).}
    \label{tab:relative_errors_nonlinear_nonhydro}
\end{table}
 
\pagebreak

\subsection{11 January 1972 Boulder Windstorm}
\label{ssec:boulder} \indent

\begin{figure}[h!]
    \centering
    \begin{subfigure}{\textwidth}
        \centering
        \includegraphics[width = 0.7\textwidth]{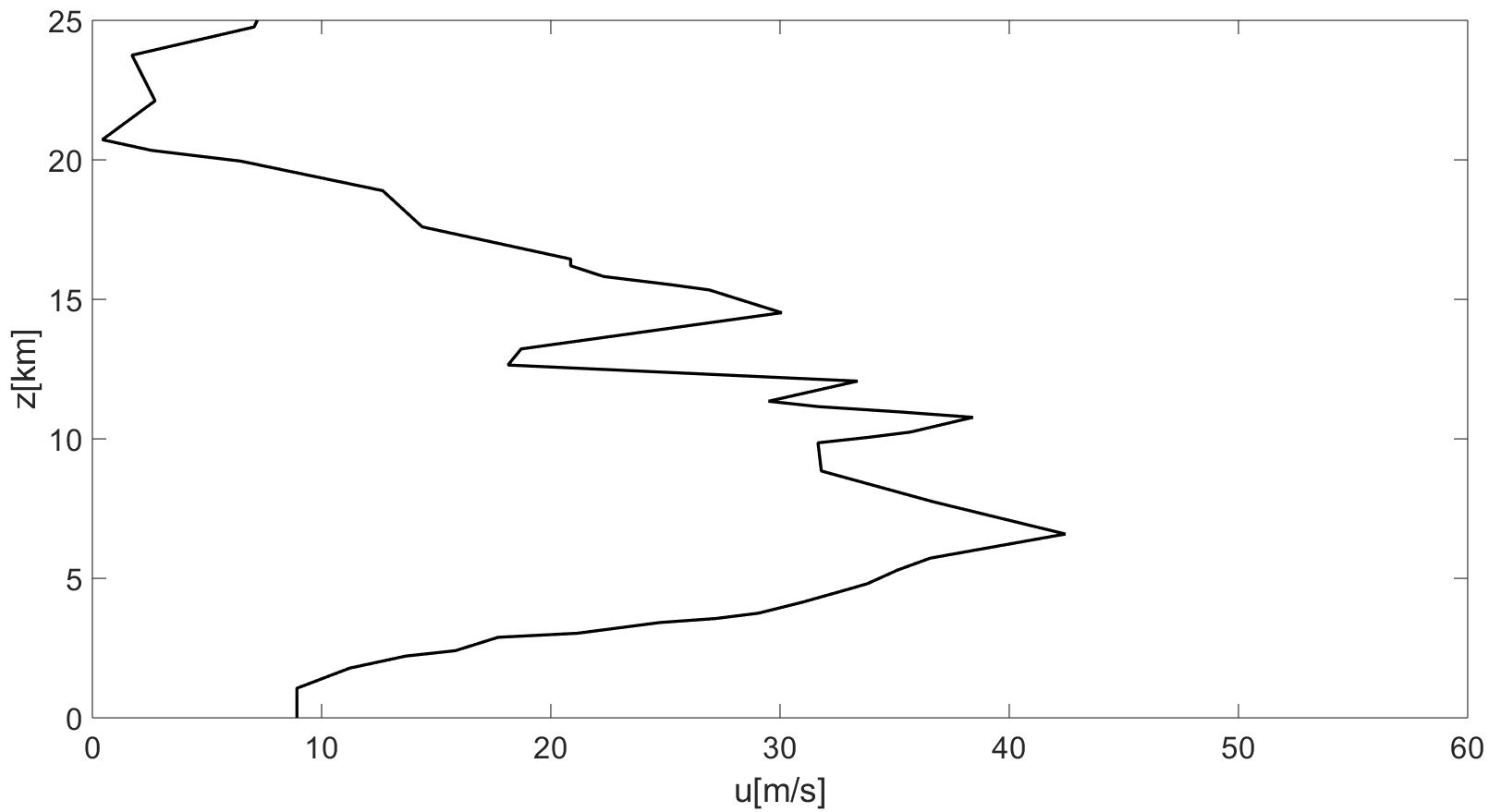}
    \end{subfigure}
    \begin{subfigure}{\textwidth}
        \centering
        \includegraphics[width = 0.7\textwidth]{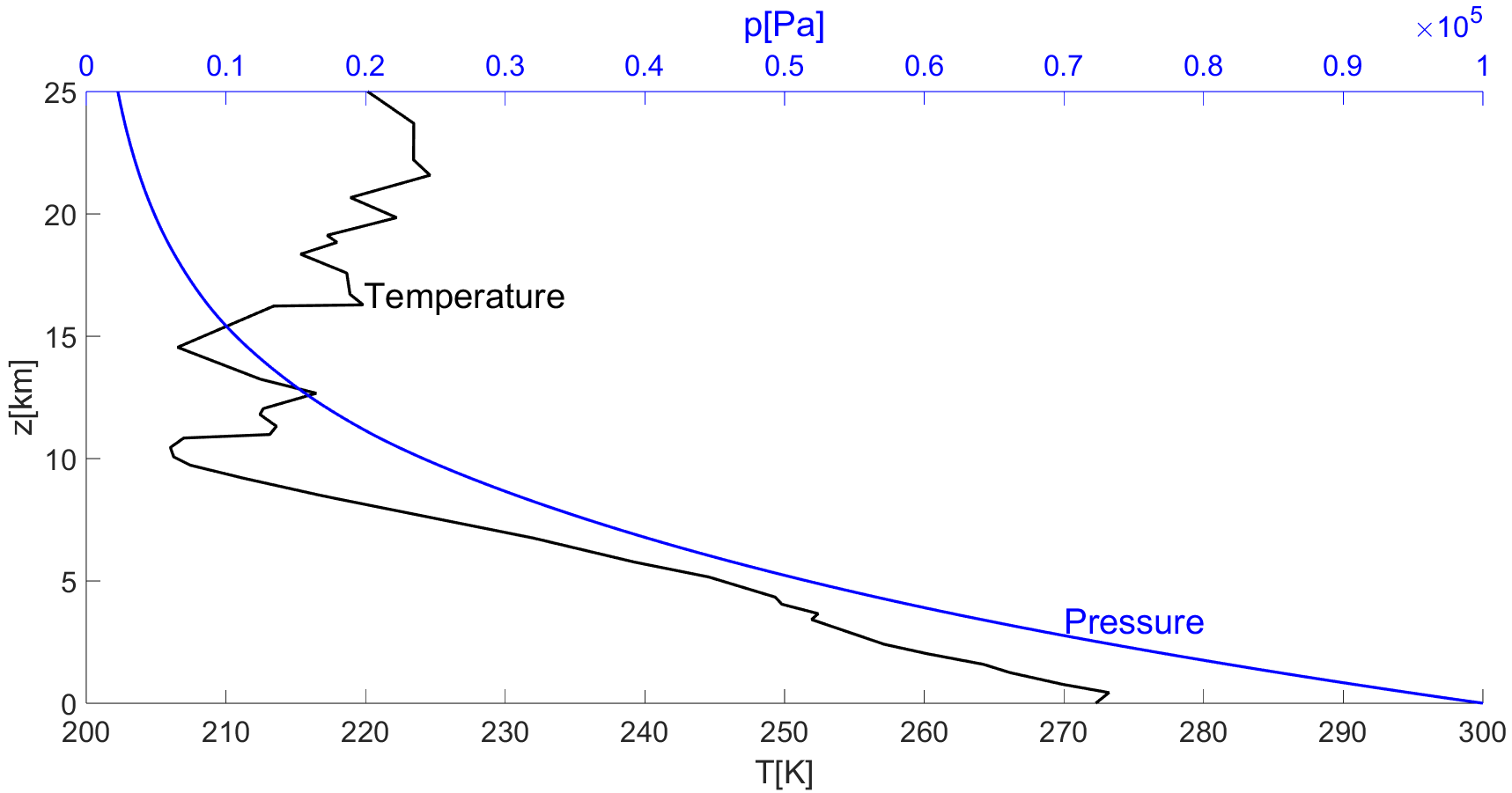}
    \end{subfigure}
    \caption{Boulder windstorm test case, initial conditions. Top: horizontal velocity. Bottom: temperature (black line) and pressure (blue line).}
    \label{fig:Boulder_initial_conditions}
\end{figure}

Next, we consider the more realistic condition of the 11 January 1972 Boulder (Colorado) windstorm benchmark \cite{doyle:2000}. This test case is particularly challenging because a complex wave-breaking response is established aloft in the lee of the mountain. The initial conditions are horizontally homogeneous and based upon the upstream measurements at 1200 UTC 11 January 1972 Grand Junction, Colorado, as shown in \cite{doyle:2000}. The initial conditions contain a critical level near \(z = \SI{21}{\kilo\meter}\) (Figure \ref{fig:Boulder_initial_conditions}), which more realistically simulates the stratospheric gravity wave breaking \cite{doyle:2000}. The pressure is computed from the hydrostatic balance, namely
\begin{equation}
    p(z) = p_{0}\exp{\left(-\frac{g}{R}\int_{0}^{z} \frac{1}{T(s)}ds\right)},
\end{equation}
with \(p_{0} = \SI[parse-numbers=false]{10^{5}}{\pascal}\). Linear interpolation is employed to evaluate both temperature and horizontal velocity.

\begin{figure}[h!]
    \centering
    \includegraphics[width=0.7\textwidth]{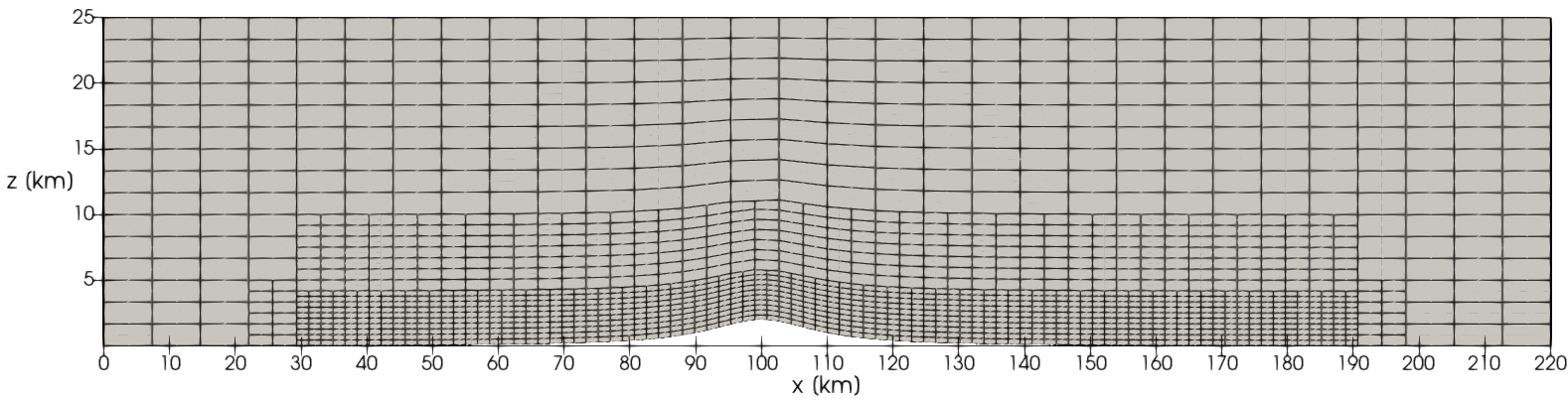}
    \caption{Boulder windstorm test case, non-conforming mesh.}
    \label{fig:Boulder_non_conforming_mesh}
\end{figure}

The bottom boundary is described by \eqref{eq:versiera_Agnesi}, with \(h_{c} = \SI{2}{\kilo\meter}, x_{c} = \SI{100}{\kilo\meter},\) and \(a_{c} = \SI{10}{\kilo\meter}\). We consider two different computational meshes: a uniform mesh composed by \(120 \times 60 = 7200\) elements, i.e. a resolution of \(\SI{458.33}{\meter}\) along the horizontal direction and of \(\SI{104.17}{\meter}\) along the vertical one, and a non-conforming mesh with 3 different levels, composed by \(N_{el} = 1524\) (Figure \ref{fig:Boulder_non_conforming_mesh}), with the finest level corresponding to the resolution of the uniform mesh.

\begin{figure}[h!]
    \centering
    \begin{subfigure}{0.9\textwidth}
        \centering
        \includegraphics[width = 0.8\textwidth]{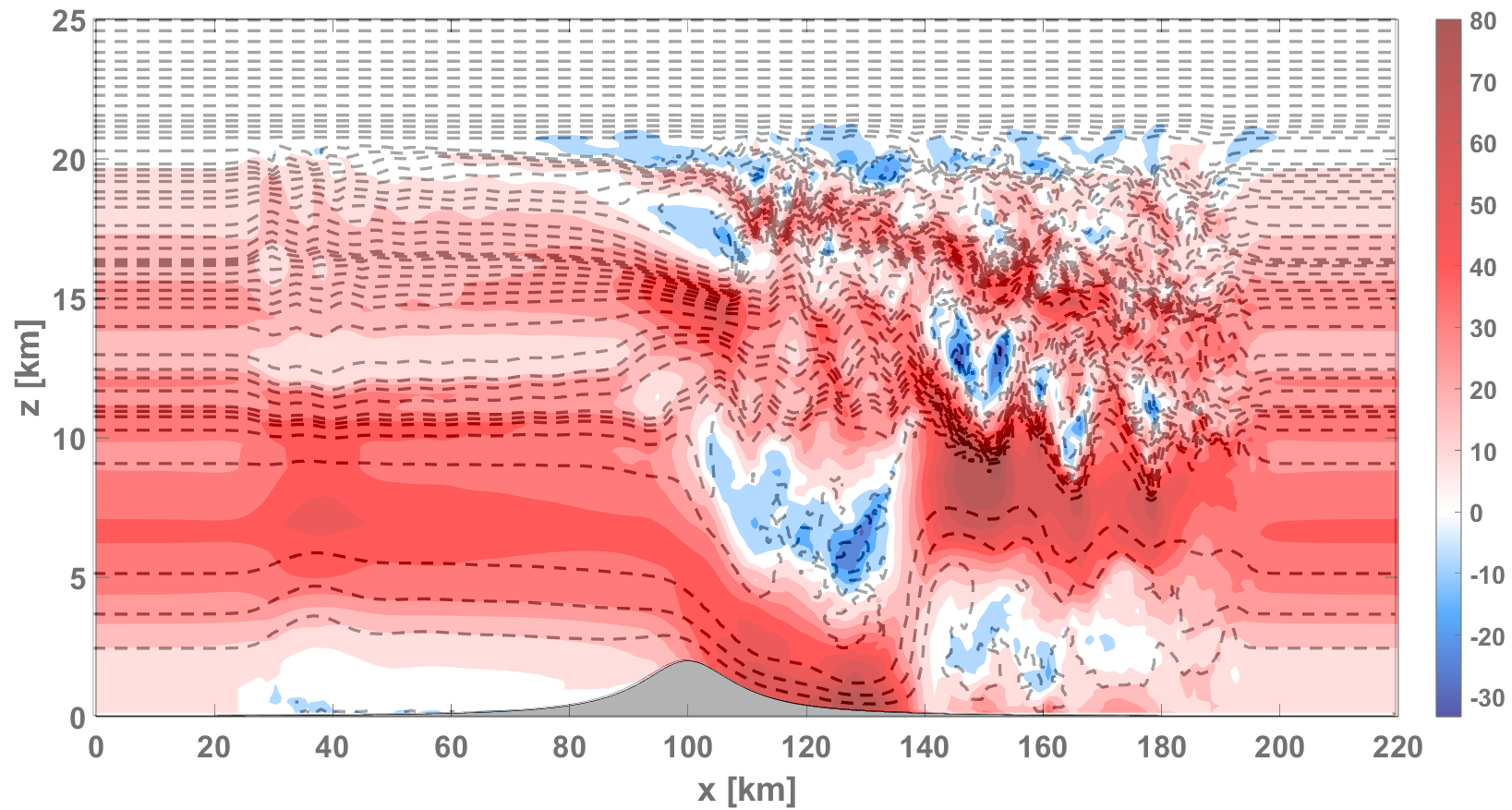}
    \end{subfigure}
    \begin{subfigure}{0.9\textwidth}
        \centering
        \includegraphics[width = 0.8\textwidth]{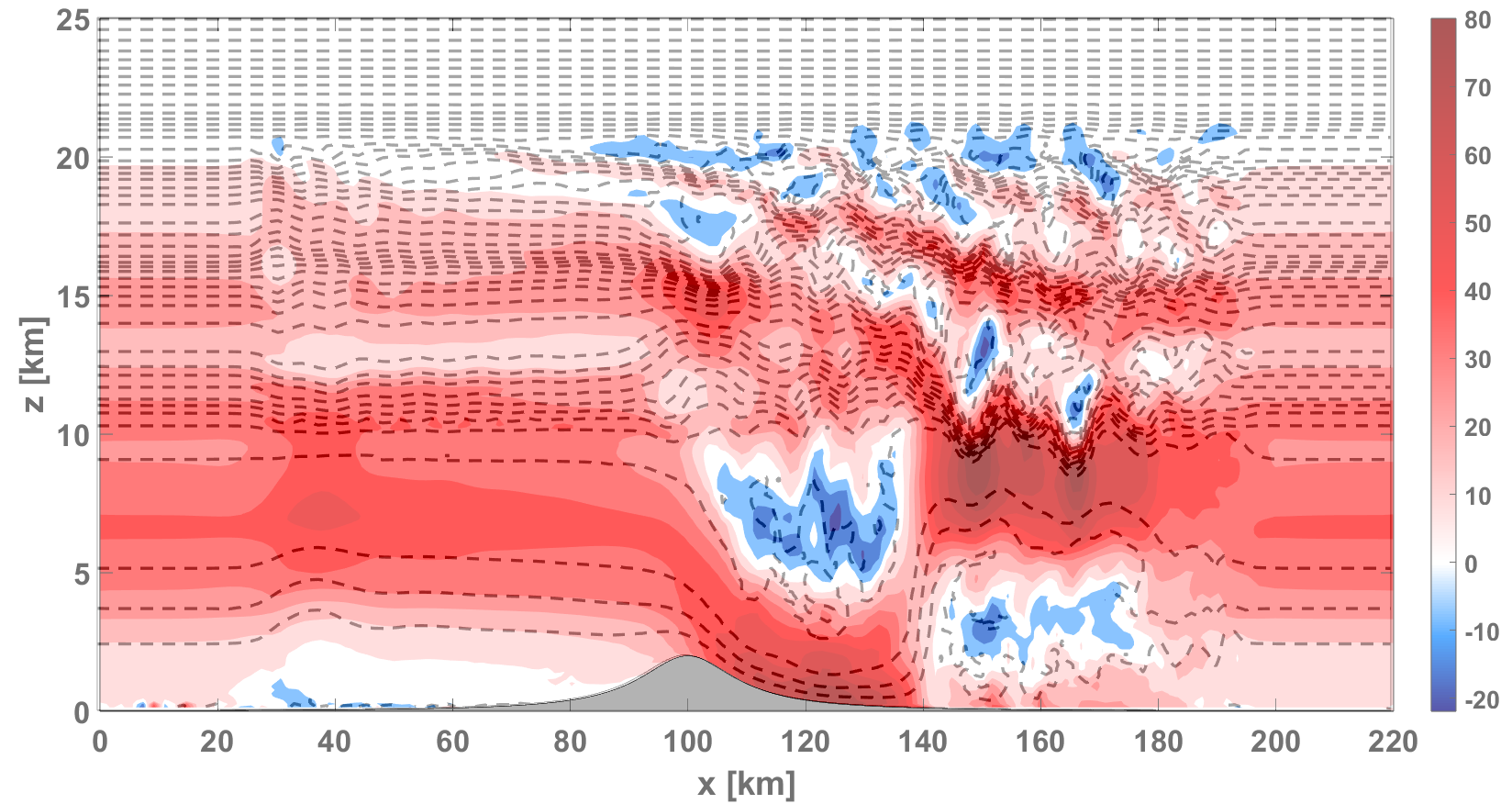}
    \end{subfigure}
    \caption{Boulder windstorm test case numerical results at \(t = T_{f} = \SI{3}{\hour}\). Top: uniform mesh. Bottom: non-conforming mesh. Horizontal velocity (colors), contours in the range \(\SI[parse-numbers=false]{[-40, 80]}{\meter\per\second}\) with a \(\SI{8}{\meter\per\second}\) interval. Potential temperature (dashed lines), contours in the range \(\SI[parse-numbers=false]{[273,650]}{\kelvin}\) with an \(\SI{8}{\kelvin}\) interval.}
    \label{fig:Boulder_contours_inviscid}
\end{figure}

The horizontal velocity and the potential temperature computed at \(t = T_{f}\) by the IMEX-DG method using a uniform mesh are in reasonable agreement with the reference results \cite{doyle:2000}, in particular for what concerns the potential temperature (Figure \ref{fig:Boulder_contours_inviscid}). Numerous regions of small-scale motion and larger high-frequency spatial structures arise with respect to the other tests using the uniform mesh, because of the lack of a subgrid eddy viscosity. The qualitative behaviour of the simulation with the uniform mesh and the one with the non-conforming mesh is in good agreement, even though visible differences arise in the deep regions of wave breaking in the stratosphere (Figure \ref{fig:Boulder_contours_inviscid}). In terms of wall-clock time, the configuration with the non-conforming mesh is about \(65\%\) computationally cheaper than the configuration with the uniform mesh (\(\SI[parse-numbers=false]{3.75 \times 10^{4}}{\second}\) vs. \(\SI[parse-numbers=false]{1.37 \times 10^{4}}{\second}\)).

\begin{figure}[h!]
    \centering
    \includegraphics[width=0.9\textwidth]{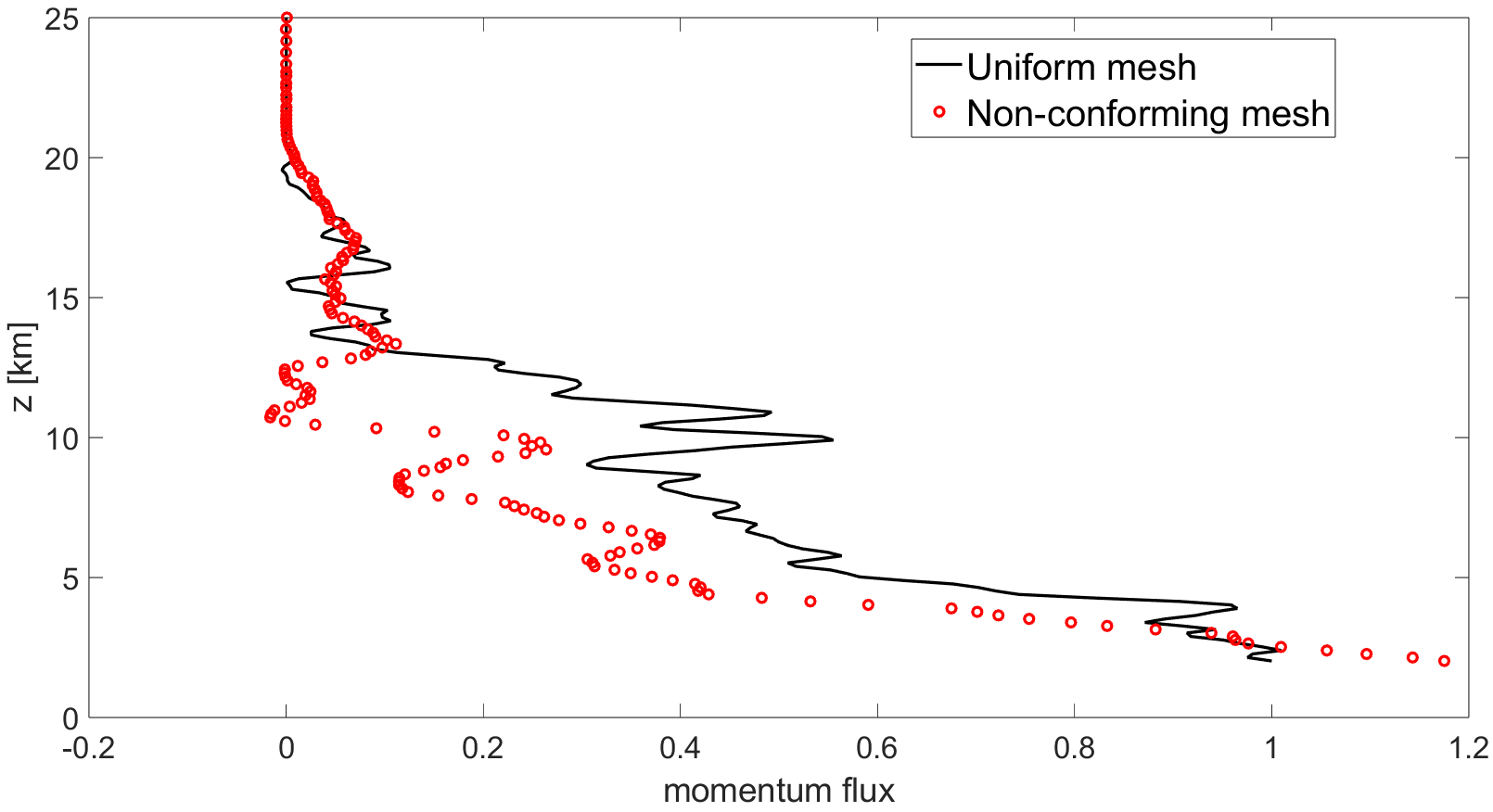}
    \caption{Boulder windstorm test case, comparison of normalized momentum flux at \(t = T_{f} = \SI{3}{\hour}\) computed using the uniform mesh (solid black line) and the non-conforming mesh (red dots).}
    \label{fig:Boulder_momentum_inviscid}
\end{figure}

Following \cite{doyle:2011}, we then compute the momentum flux \eqref{eq:momentum_flux_def} using the mean value of \(u\) and \(w\) to compute \(u'\) and \(w'\). A comparison at final time of the vertical flux of horizontal momentum \eqref{eq:momentum_flux_def} normalized by its values at the surface obtained with the uniform mesh displays a reasonable agreement between the two simulations, especially for \(z\) above \(\SI{12}{\kilo\meter}\) (Figure \ref{fig:Boulder_momentum_inviscid}). The discrepancy in the vertical region between \(z = \SI{7}{\kilo\meter}\) and \(z = \SI{12}{\kilo\meter}\) is probably due to the development of small-scale features and to the lack of subgrid eddy viscosity parametrization, as already discussed for the contour plots.

\pagebreak

Next, we repeat the simulations for this test case including a simplified model for turbulent vertical diffusion for NWP applications, originally proposed in \cite{louis:1979} and also discussed in \cite{benard:2000, bonaventura:2014, girard:1990}. As commonly done in numerical models for atmospheric physics, we resort to an operator splitting approach. The diffusion model is treated with the implicit part of the IMEX method, which corresponds to the TR-BDF2 scheme \cite{hosea:1996, orlando:2022}. The non-linear diffusivity \(\kappa\) has the form
\begin{equation}\label{eq:turbulent_diffusion}
    \kappa\left(\frac{\partial u}{\partial z}, \frac{\partial\theta}{\partial z}\right) = l^{2}\left|\frac{\partial u}{\partial z}\right|F\left(Ri\right).
\end{equation}
Here, \(l\) is a mixing length and \(Ri\) is the Richardson number given by
\begin{equation}
    Ri = \frac{g}{\theta_{0}}\frac{\frac{\partial\theta}{\partial z}}{\left|\frac{\partial u}{\partial z}\right|^{2}},
\end{equation}
with \(\theta_{0}\) denoting a reference temperature. Finally, the function \(F\left(Ri\right)\) is defined as
\begin{equation}
    F\left(Ri\right) = \left(1 + b\left|Ri\right|\right)^{\beta},
\end{equation}
where
\begin{equation}
    \begin{cases}
        \beta = -2, b = 5 \qquad &\text{if } Ri > 0 \\
        \beta = \frac{1}{2}, b = 20 \qquad &\text{if } Ri < 0.
    \end{cases}
\end{equation} 
We consider the uniform mesh with \(120 \times 60 = 7200\) elements and a coarse non-conforming mesh with \(N_{el} = 1524\) elements already employed for the inviscid case. In addition, we consider a fine non-conforming mesh with three different refinement levels and \(N_{el} = 6324\) elements. The fine resolution around the orography is of \(\SI{229.17}{\meter}\) along the horizontal direction and of \(\SI{52.08}{\meter}\) along the vertical one. We use a time step \(\Delta t = \SI{0.375}{\second}\), corresponding to a maximum acoustic Courant number \(C \approx 0.79\) and a maximum advective Courant number \(C_{u} \approx 0.23\). Finally, we take \(l = \SI{100}{\meter}\) and \(\theta_{0} = \SI{273}{\kelvin}\) in \eqref{eq:turbulent_diffusion}.

\begin{figure}[h!]
    \centering
    \includegraphics[width=0.9\textwidth]{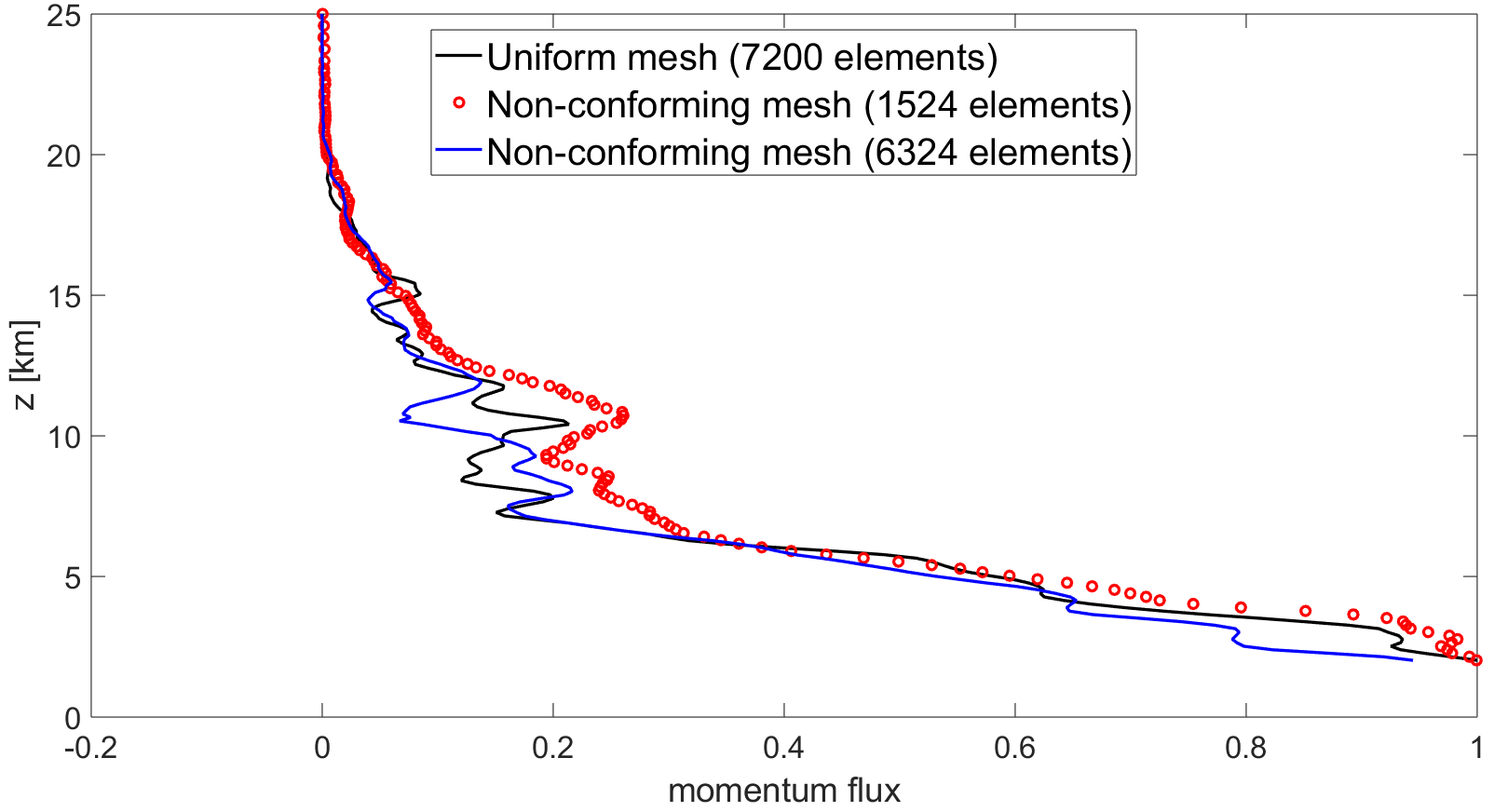}
    \caption{Boulder windstorm test case, comparison of normalized momentum flux at \(t = T_{f} = \SI{3}{\hour}\) between the uniform mesh (solid black line), the fine (blue solid line), and the coarse (red dots) non-conforming meshes.}
    \label{fig:Boulder_momentum_turbulent}
\end{figure}

\begin{figure}[h!]
    \centering
    \begin{subfigure}{0.9\textwidth}
        \centering
        \includegraphics[width = 0.8\textwidth]{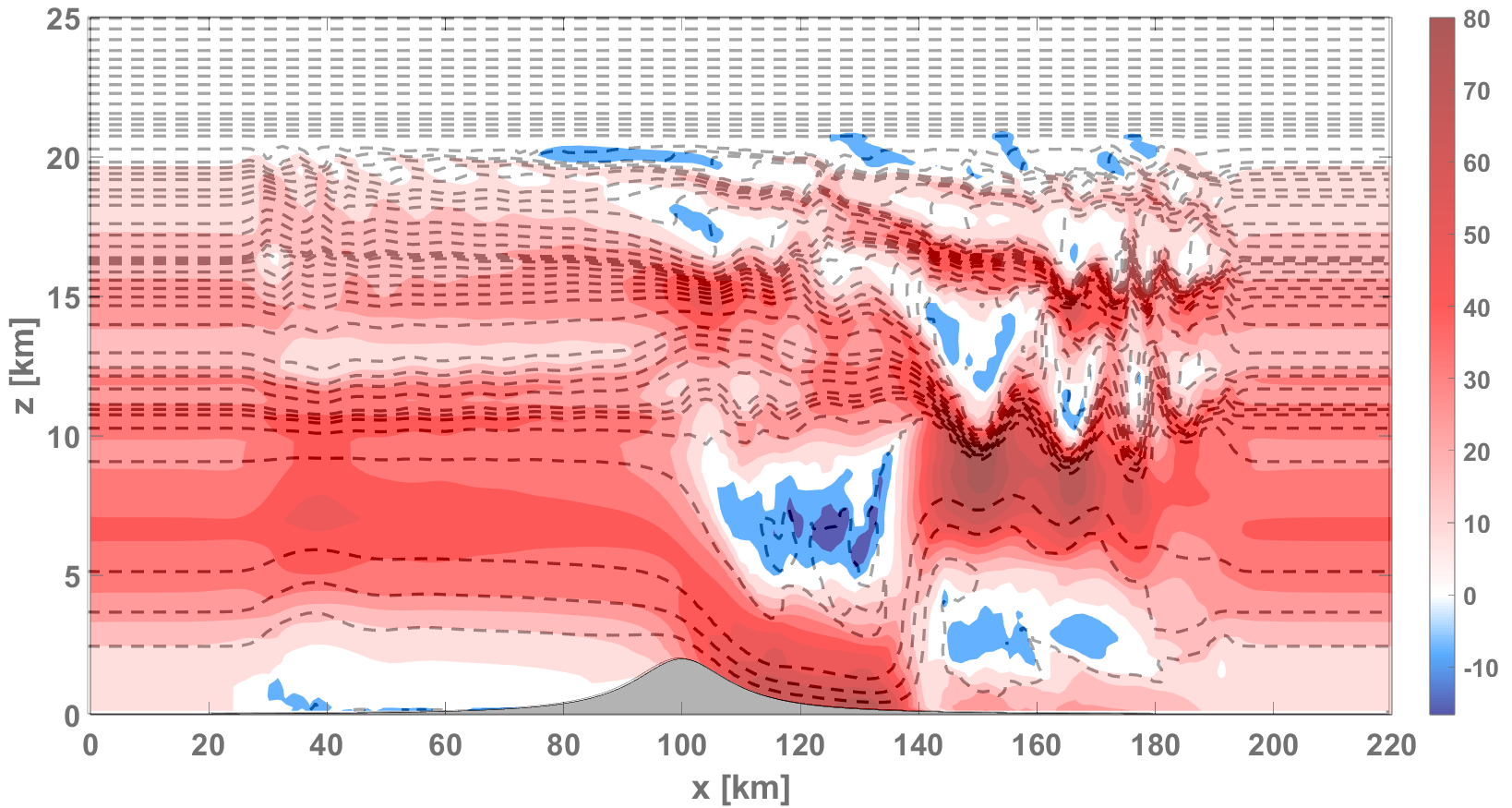}
    \end{subfigure}
    \begin{subfigure}{0.9\textwidth}
        \centering
        \includegraphics[width = 0.8\textwidth]{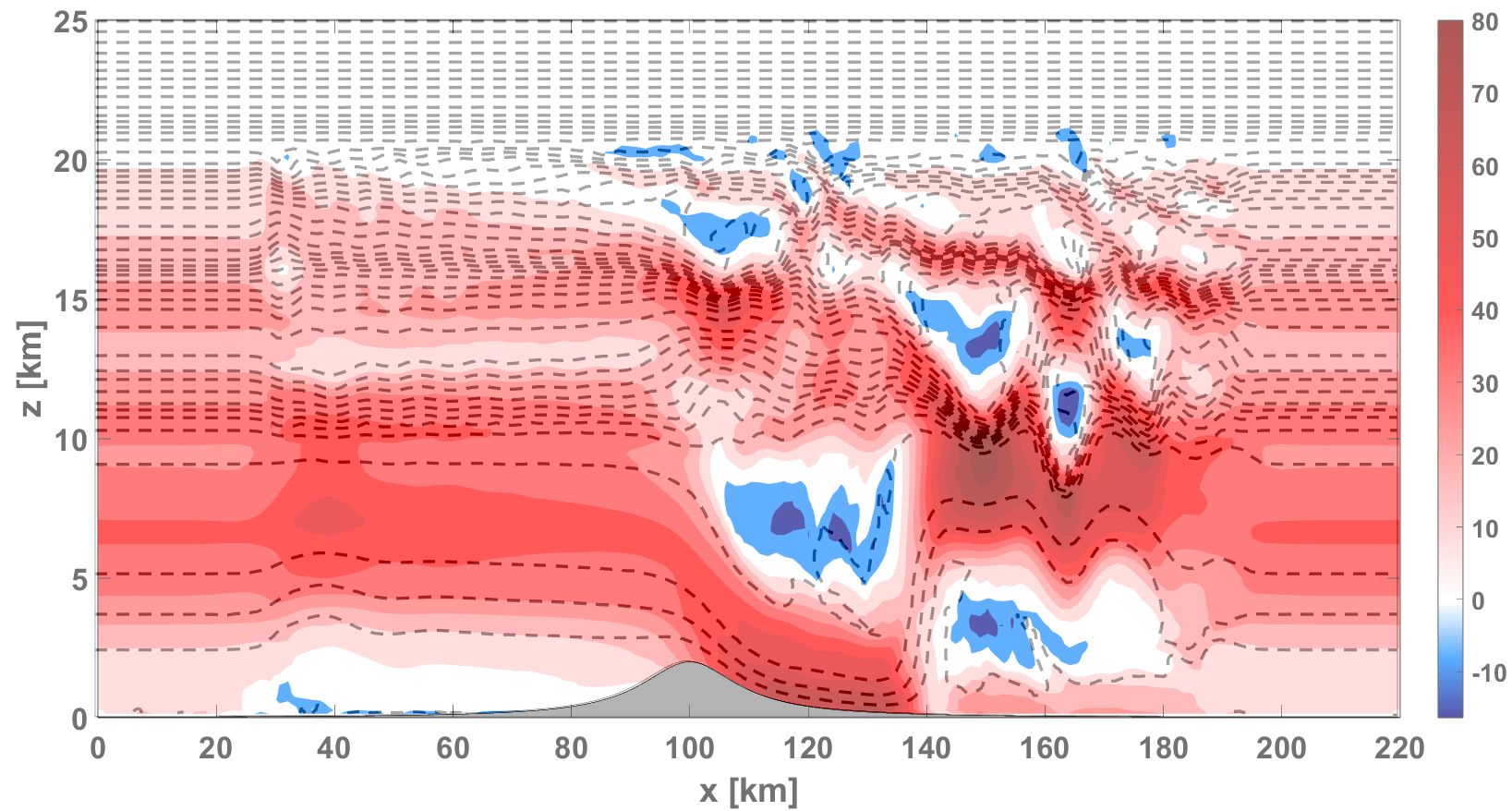}
    \end{subfigure}
    \begin{subfigure}{0.9\textwidth}
        \centering
        \includegraphics[width = 0.8\textwidth]{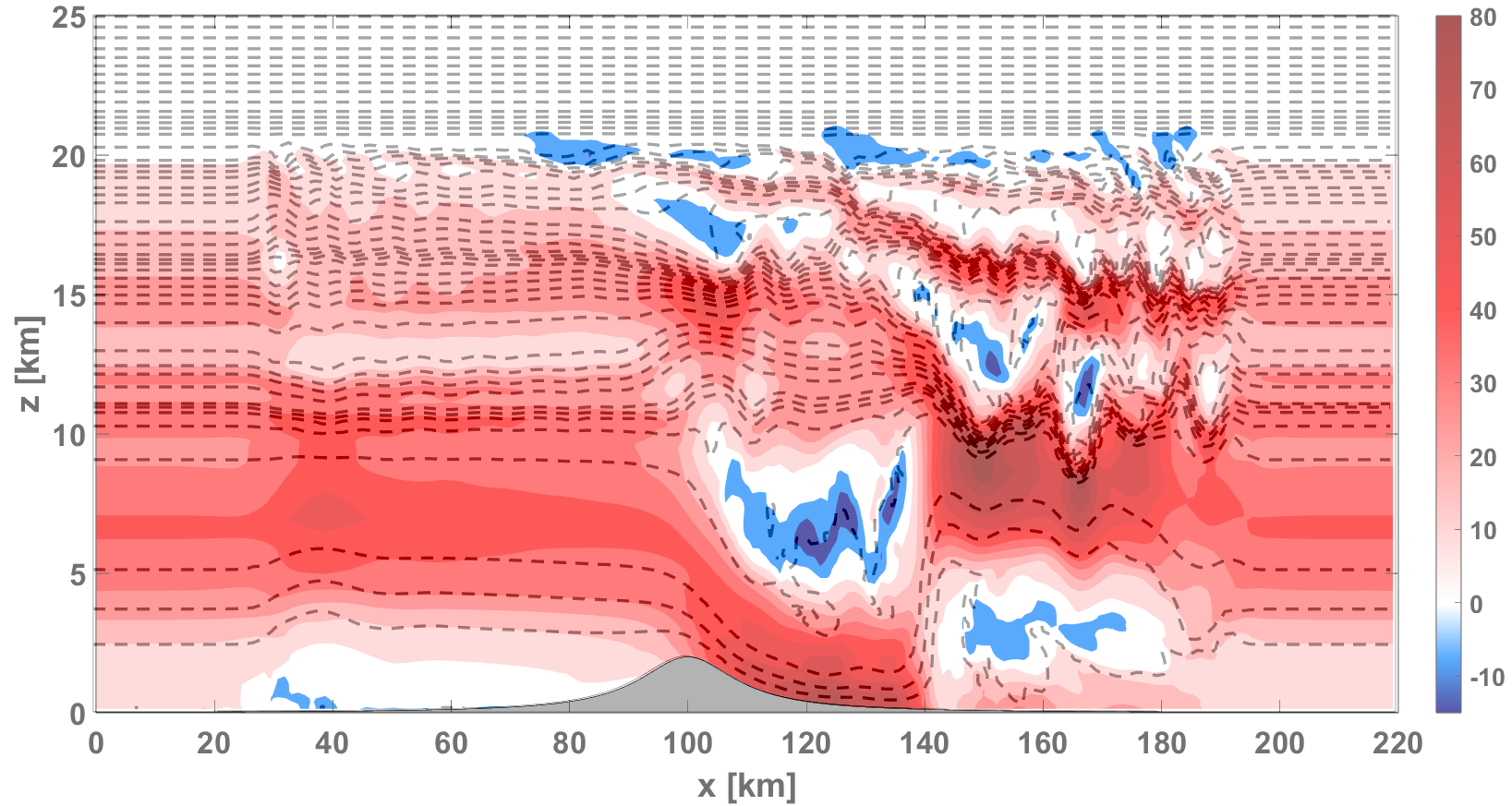}
    \end{subfigure}
    \caption{Boulder windstorm test case with turbulent vertical diffusion. Top: uniform mesh. Middle: coarse non-conforming mesh. Bottom: fine non-conforming mesh. Horizontal velocity (colors), contours in the range \(\SI[parse-numbers=false]{[-40, 80]}{\meter\per\second}\) with a \(\SI{8}{\meter\per\second}\) interval. Potential temperature (dashed lines), contours in the range \(\SI[parse-numbers=false]{[273,650]}{\kelvin}\) with an \(\SI{8}{\kelvin}\) interval.}
    \label{fig:Boulder_contours_turbulent}
\end{figure}

At \(t = T_{f}\), numerical solutions computed using the uniform mesh and the non-conforming mesh are in good agreement for both the horizontal velocity and the potential temperature, in particular for the finest non-conforming mesh (Figure \ref{fig:Boulder_contours_turbulent}). In terms of wall-clock time, a computational time saving of around \(50\%\) is achieved with the coarse non-conforming mesh (bold numbers in Table \ref{tab:wall_clock_Boulder}), while performance is less optimal for the fine non-conforming mesh (see the discussion above in Section \ref{ssec:hydrostatic}). Finally, a comparison at \(t = T_{f}\) of the computed momentum flux \eqref{eq:momentum_flux_def} normalized by its values at the surface obtained with the uniform mesh suggests that the results of the uniform mesh are approached as long as the resolution of the non-conforming mesh increases (Figure \ref{fig:Boulder_momentum_turbulent}).

\begin{table}[h!]
    \centering
    \footnotesize
    \begin{tabularx}{0.6\columnwidth}{rXXrr}
	\toprule
        \(N_{el}\) & \(\Delta x[\SI{}{\meter}]\) & \(\Delta z[\SI{}{\meter}]\) & WT\([\SI{}{\second}]\) & Speed-up \\
	\midrule
	7200 (uniform) & 458.33 & 104.17 & \textbf{119000} & \\
	\midrule
        1524 (non-conforming) & 458.33 & 104.17 & \textbf{51900} & \textbf{2.3} \\
	\midrule
	6324 (non-conforming) & 229.17 & 52.08 & 160000 & \\
	\bottomrule
    \end{tabularx}
    \caption{Boulder windstorm test case with turbulent vertical diffusion: horizontal resolution \(\Delta x\), vertical resolution \(\Delta z\), and wall-clock times (WT) for the uniform mesh and the non-conforming meshes. The speed-up is computed considering the same maximum spatial resolution, i.e. comparing the wall-clock time of the finest uniform mesh and the wall-clock time of the non-conforming mesh (bold WT, see also main text for further details).}
    \label{tab:wall_clock_Boulder}
\end{table}

\FloatBarrier

\subsection{T-REX Mountain-Wave}
\label{ssec:trex} \indent

\begin{figure}[h!]
    \centering
    \includegraphics[width = 0.7\textwidth]{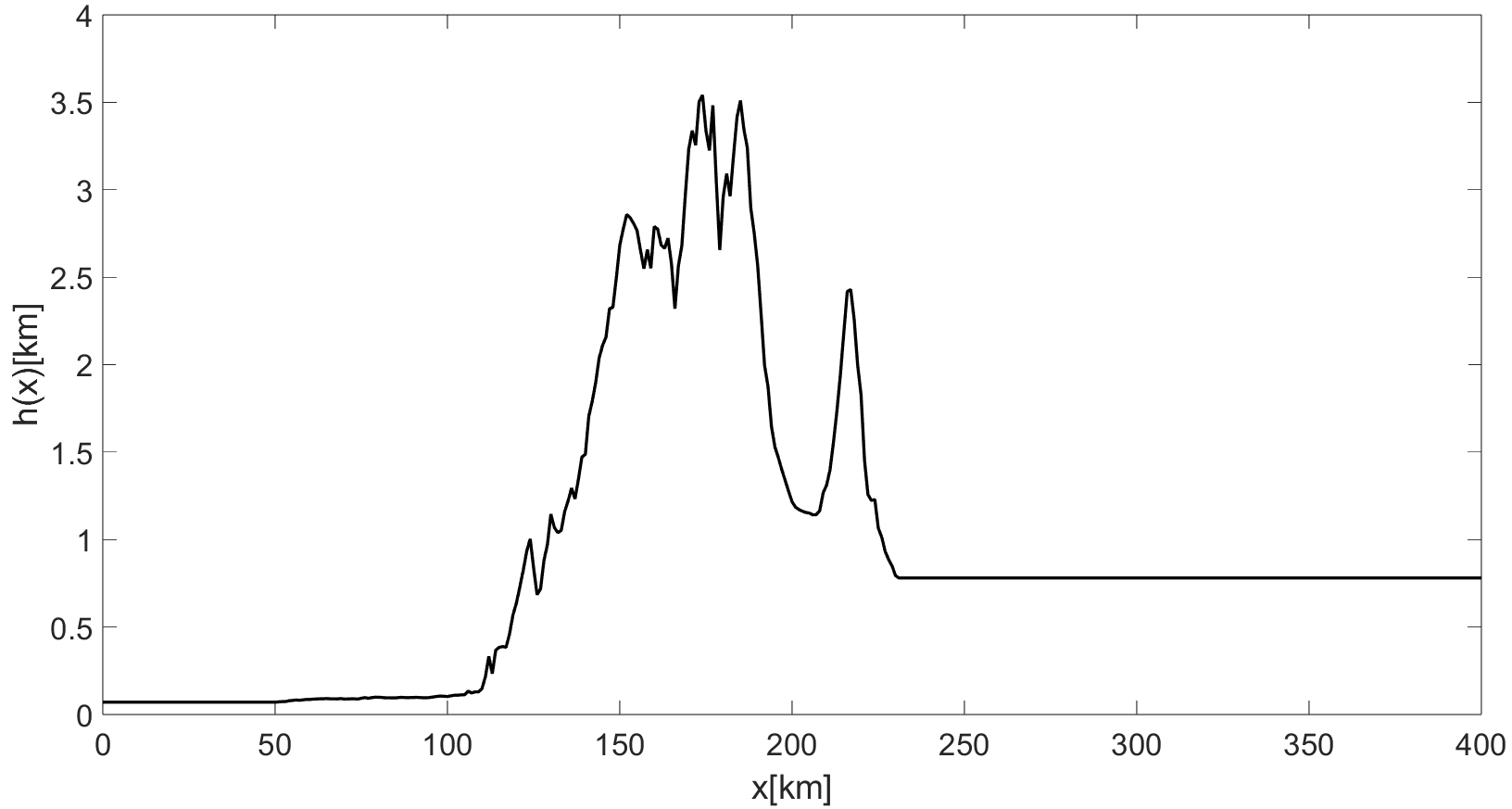}
    \caption{T-REX mountain-wave test, Sierra profile.}
    \label{fig:TREX_profile}
\end{figure}

\begin{figure}[h!]
    \centering
    \begin{subfigure}{\textwidth}
        \centering
        \includegraphics[width = 0.7\textwidth]{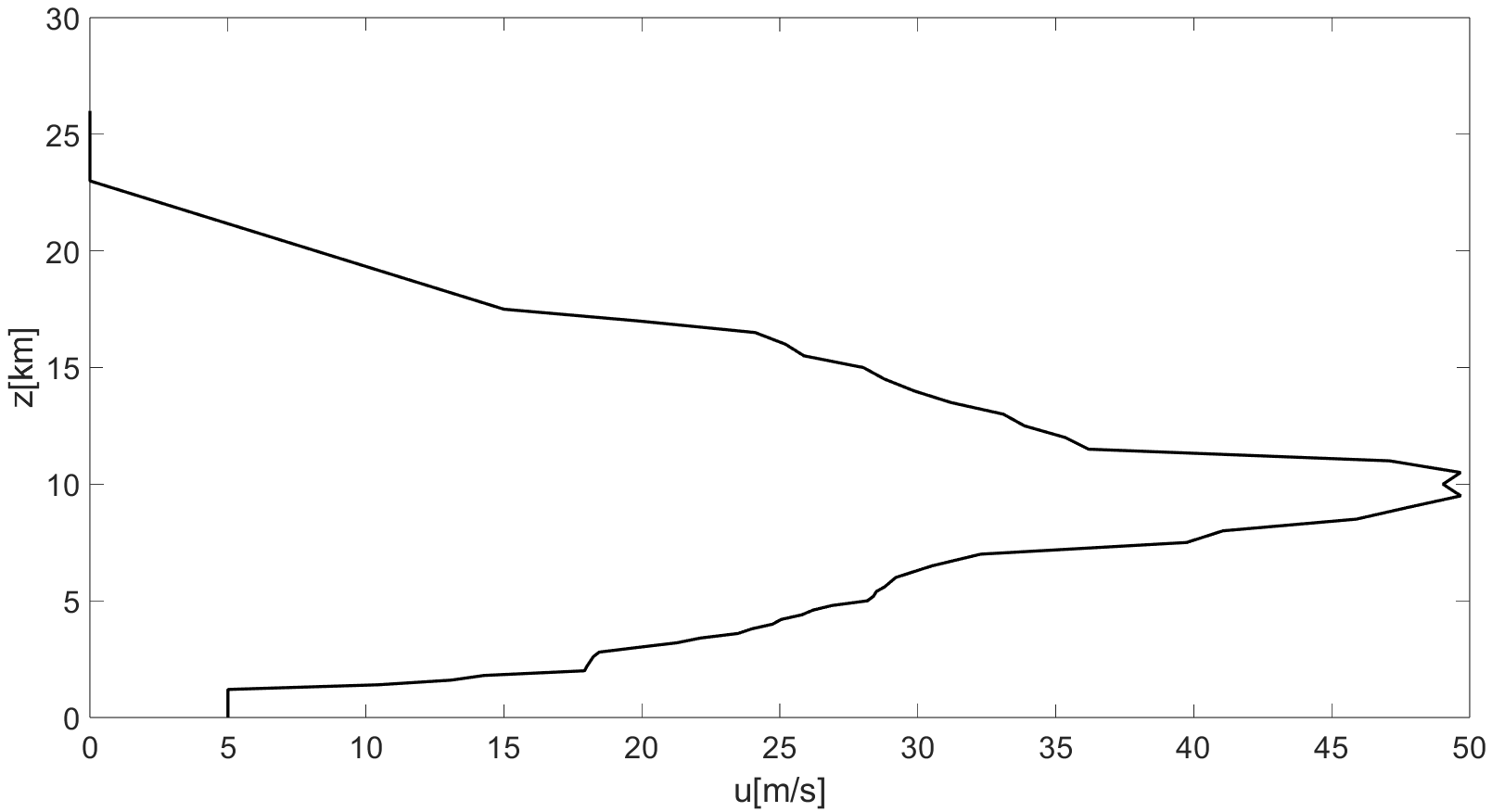}
    \end{subfigure}
    \begin{subfigure}{\textwidth}
        \centering
        \includegraphics[width = 0.7\textwidth]{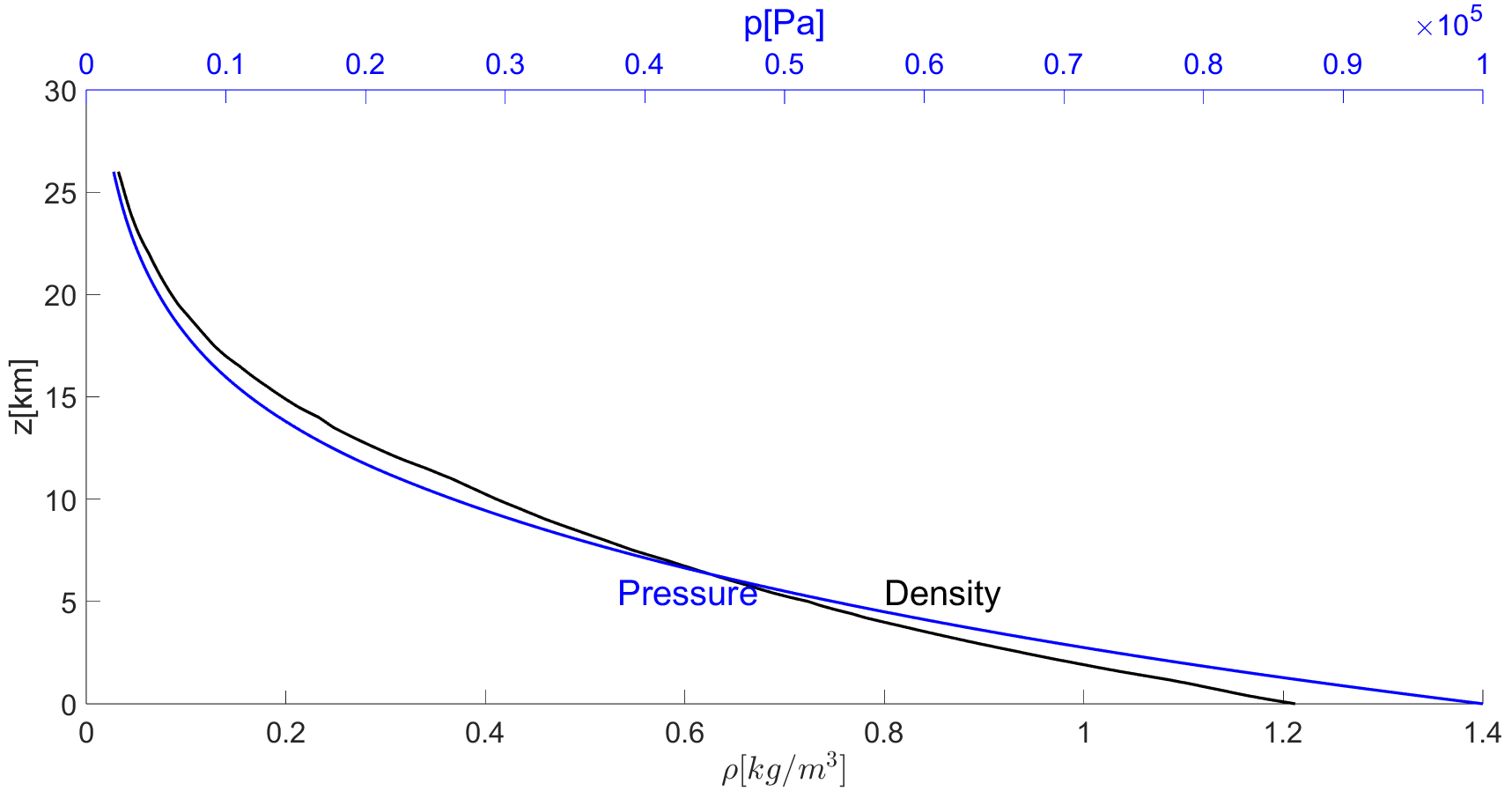}
    \end{subfigure}
    \caption{T-REX mountain-wave test case, initial conditions. Top: horizontal velocity. Bottom: density (black line) and pressure (blue line).}
    \label{fig:TREX_initial_conditions}
\end{figure}

\begin{figure}[h!]
    \centering
    \includegraphics[width=0.9\textwidth]{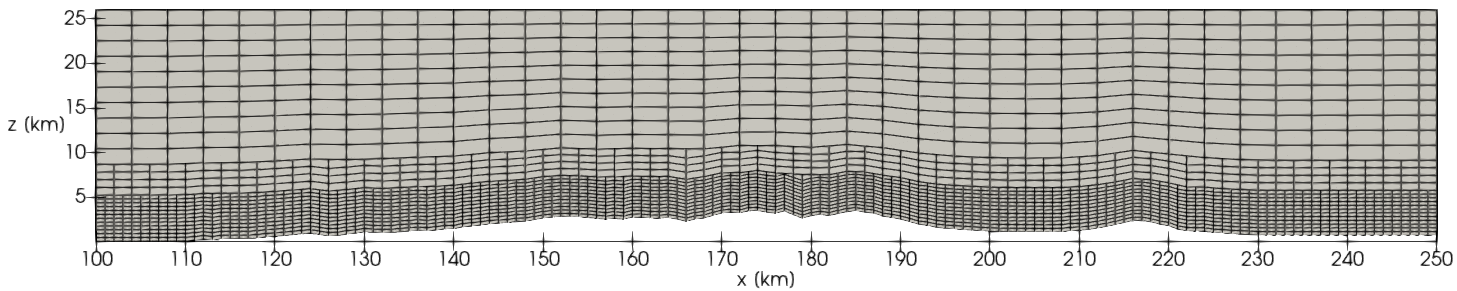}
    \caption{T-REX mountain wave, non-conforming mesh.}
    \label{fig:TREX_non_conforming_mesh}
\end{figure}

Next, we consider simulations of a flow over a steep real orography \cite{doyle:2011, kuhnlein:2013}, as shown in Figure \ref{fig:TREX_profile}. The initial state is horizontally homogeneous and it is based on conditions during Intensive Observation Period (IOP) 6 of the Terrain-Induced Rotor Experiment (T-REX) \cite{doyle:2011}, as reported in Figure \ref{fig:TREX_initial_conditions}. We consider a DG spatial discretization using degree \(r = 2\) polynomials and three computational meshes: a uniform mesh composed by \(400 \times 60=24000\) elements, corresponding to a resolution of \(\SI{500}{\meter}\) along the horizontal direction and of \(\SI{216.66}{\meter}\) along the vertical direction, and two non-conforming meshes. The coarsest non-conforming mesh consists of three different levels and \(N_{el} = 5298\) elements, whereas the finest non-conforming mesh is obtained with a global refinement of the coarsest non-conforming mesh, with \(N_{el} = 21792\) elements (Figure \ref{fig:TREX_non_conforming_mesh}). The finest level of the coarsest non-conforming mesh corresponds to the resolution of the uniform mesh. Hence, the fine resolution around the orography for the finest non-conforming mesh is \(\SI{250}{\meter}\) along the horizontal direction and \(\SI{108.33}{\meter}\) along the vertical direction. We take \(l = \SI{100}{\meter}\) and \(\theta_{0} = \SI{273}{\kelvin}\) in \eqref{eq:turbulent_diffusion}. The vertical turbulent diffusion model is necessary to obtain a stable numerical solution.

\begin{figure}[h!]
    \centering
    \begin{subfigure}{0.9\textwidth}
        \centering
        \includegraphics[width = 0.8\textwidth]{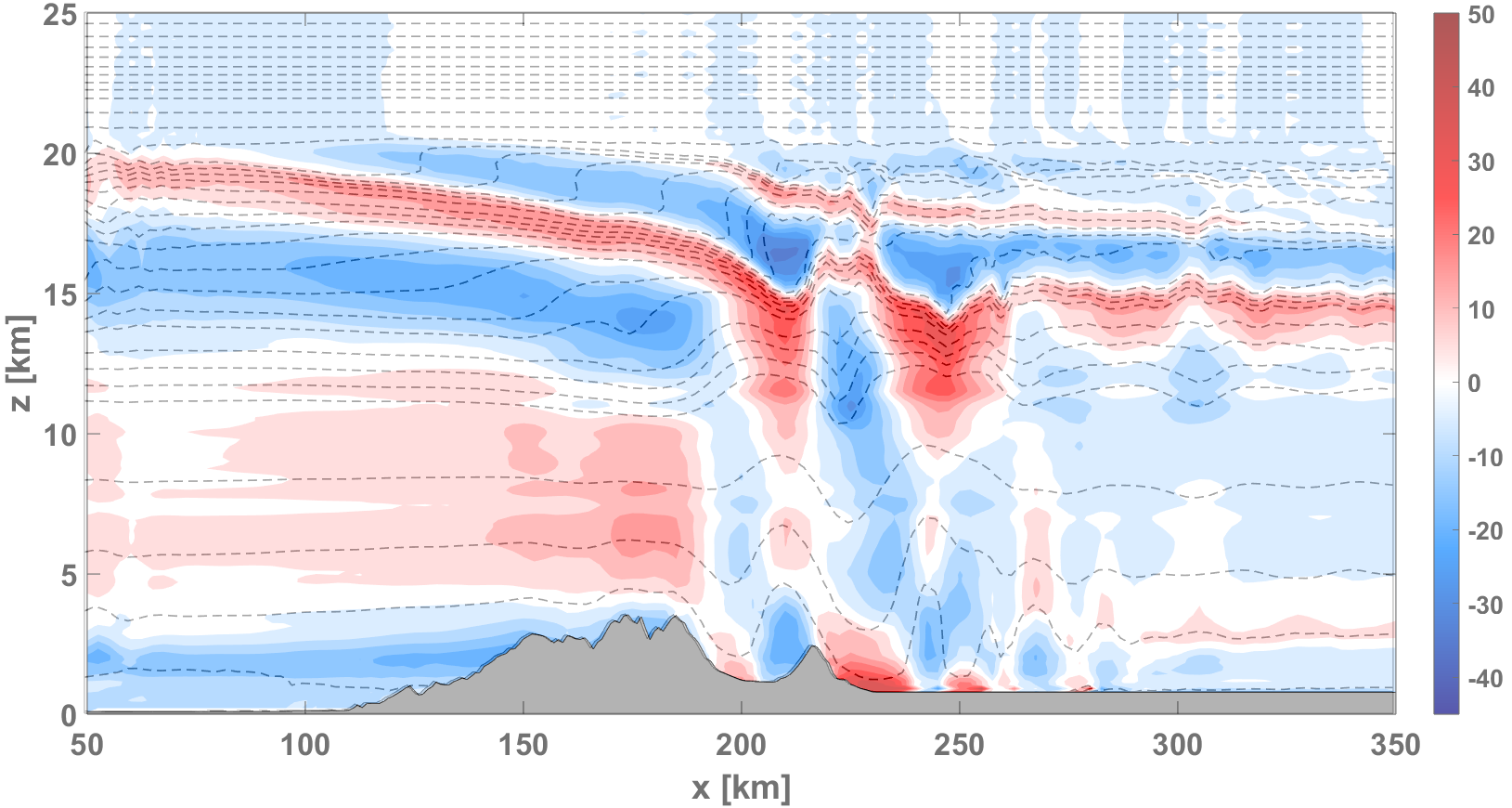}
    \end{subfigure}
    \begin{subfigure}{0.9\textwidth}
        \centering
        \includegraphics[width = 0.8\textwidth]{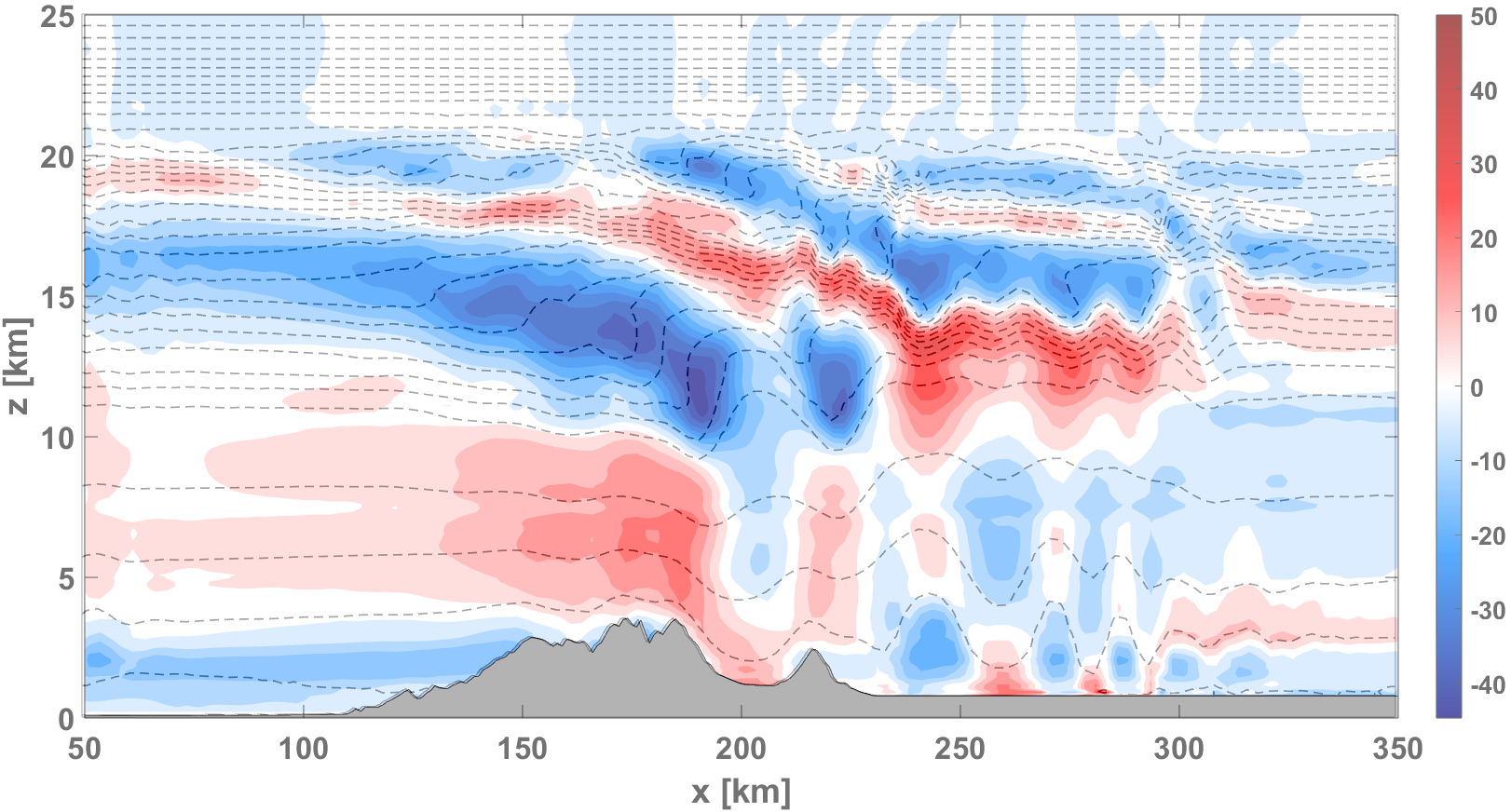}
    \end{subfigure}
    \begin{subfigure}{0.9\textwidth}
        \centering
        \includegraphics[width = 0.8\textwidth]{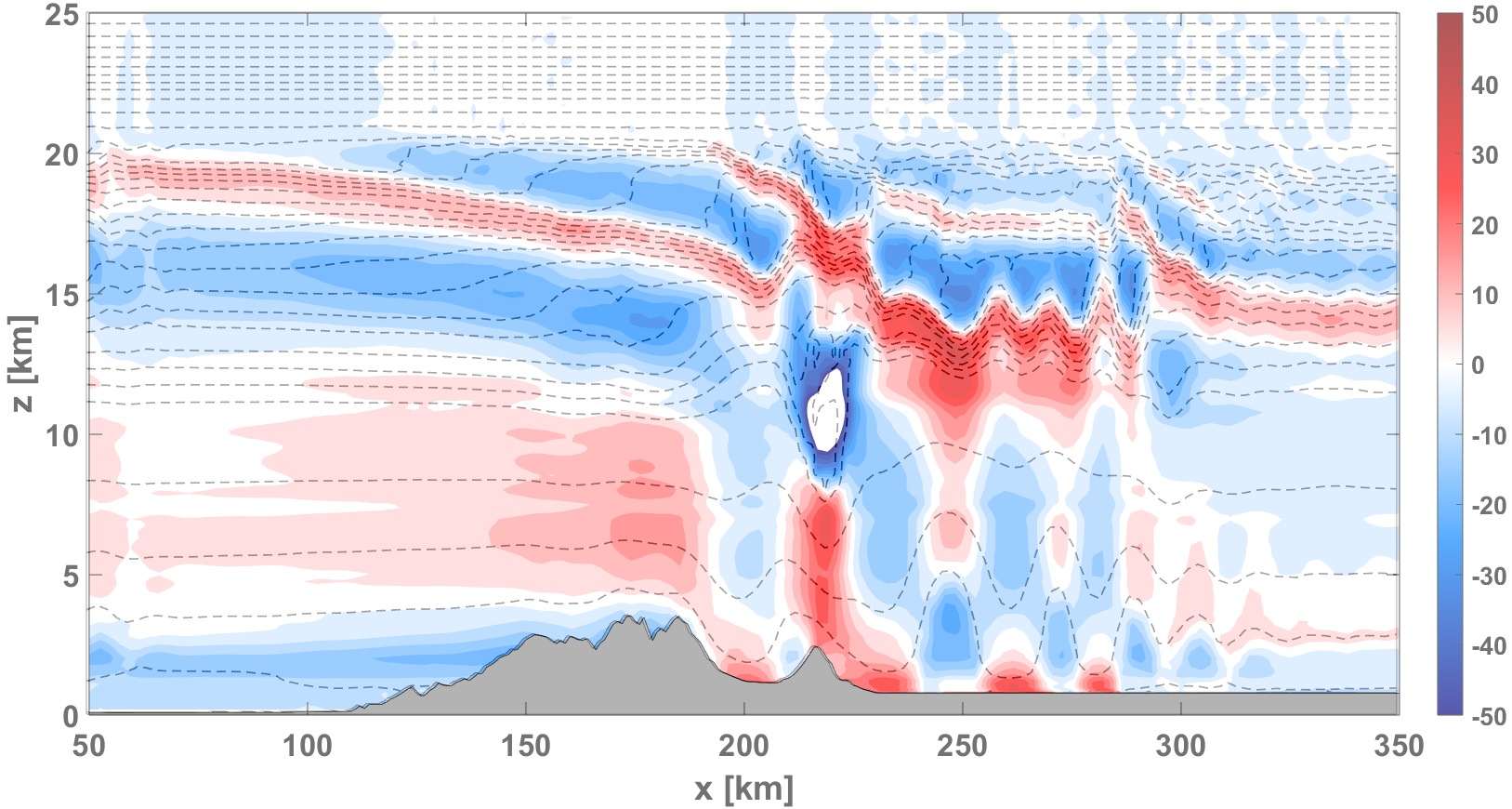}
    \end{subfigure}
    \caption{T-REX mountain wave test case at \(T_{f} = \SI{4}{\hour}\). Top: uniform mesh. Middle: coarsest non-conforming mesh. Bottom: finest non-conforming mesh. Horizontal velocity perturbation (colors), contours in the range\(\SI[parse-numbers=false]{[-50,50]}{\meter\per\second}\) with a \(\SI{2.5}{\meter\per\second}\) interval. Potential temperature (dashed lines), contours in the range \(\SI[parse-numbers=false]{[273, 650]}{\kelvin}\) with a \(\SI{10}{\kelvin}\) interval.}
    \label{fig:TREX_contours}
\end{figure}

Both in the horizontal velocity and in the potential temperature variables, the IMEX-DG numerical solutions at \(t = T_{f}\) display reasonable agreement between results obtained using the uniform mesh and the non-conforming meshes, showing the robustness of the proposed approach based on non-conforming meshes also in the case of a realistic, steep orography (Figure \ref{fig:TREX_contours}). Some differences arise in the structures of the horizontal velocity, but, unlike the previous test case, in this benchmark \cite{doyle:2011} there is low predictability of key characteristics such as the strength of downslope winds or the stratospheric wave breaking. Moreover, the change in the resolution of the topography has been shown to modify the representation of mountain wave-driven middle atmosphere processes \cite{kanehama:2019}. The contour plots show overall a reasonable agreement with those reported in \cite{doyle:2011}. While we have employed the same range values adopted in \cite{doyle:2011}, one can easily notice that a lower minimum value of the velocity around \(x \approx \SI{220}{\kilo\meter}\) and \(z \approx \SI{11}{\kilo\meter}\) is achieved for the finest non-conforming mesh. This is likely due to the use of a high-order method with low numerical dissipation and to the increased resolution. For the sake of completeness, we have also run a simulation up to \(t = \SI{5}{\hour}\) and no numerical instability arises.

\begin{figure}[h!]
    \centering
    \includegraphics[width=0.9\textwidth]{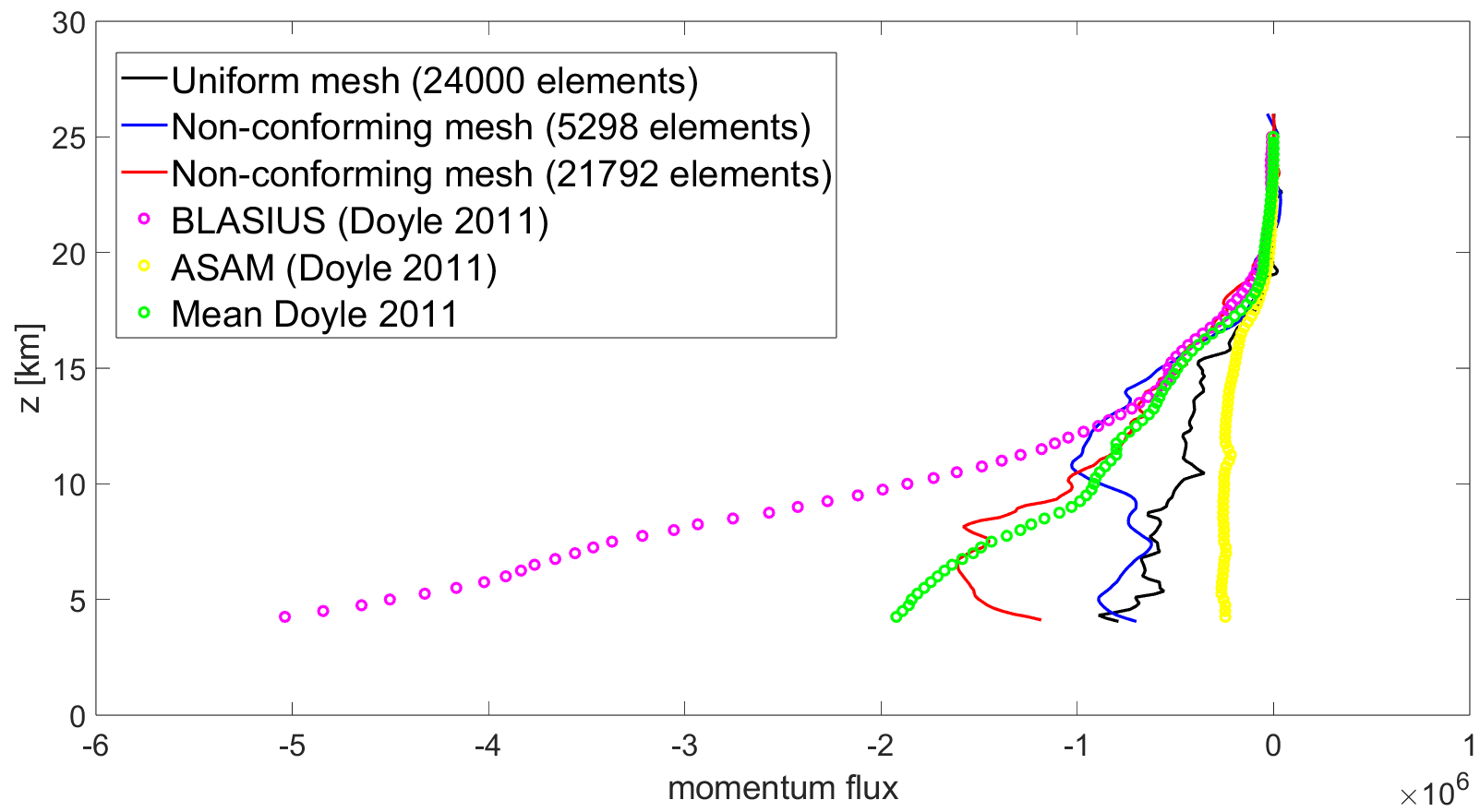}
    \caption{T-REX mountain wave test case, comparison of momentum flux at \(t = T_{f} = \SI{4}{\hour}\) between the uniform mesh (solid black line), the finest non-conforming mesh (solid blue line), and the coarsest non-conforming mesh (solid red line). Results obtained in \cite{doyle:2011} are also reported. BLASIUS model (magenta dots), ASAM model (yellow dots), mean of all the models (green dots).}
    \label{fig:TREX_momentum}
\end{figure}

A far-field comparison of the momentum flux \eqref{eq:momentum_flux_def} confirms the low predictability of large-scale orographic features for this test case (Figure \ref{fig:TREX_momentum}). On the other hand, one can easily notice that the momentum flux profiles shown in Figure \ref{fig:TREX_momentum} yield values of the same order of magnitude as those obtained with the models compared in \cite{doyle:2011}. More specifically, the BLASIUS model employed in \cite{doyle:2011} predicts the lowest values, while the ASAM model predicts the highest ones. The values obtained in our framework, especially those established with the finest non-conforming mesh, are close to the mean values of all the models compared in \cite{doyle:2011}. Similarly to Section \ref{ssec:boulder}, a computational time saving of around \(25\%\) is achieved with the coarse non-conforming mesh (bold numbers in Table \ref{tab:wall_clock_TREX}), while performance is less optimal for the fine non-conforming mesh.

\begin{table}[h!]
    \centering
    \footnotesize
    \begin{tabularx}{0.6\columnwidth}{rXXrr}
	\toprule
        \(N_{el}\) & \(\Delta x[\SI{}{\meter}]\) & \(\Delta z[\SI{}{\meter}]\) & WT\([\SI{}{\second}]\) & Speed-up \\
	\midrule
	24000 (uniform) & 500 & 216.66 & \textbf{10900} & \\
	\midrule
        5298 (non-conforming) & 500 & 216.66 & \textbf{8390} & \textbf{1.3} \\
	\midrule
	21792 (non-conforming) & 250 & 108.33 & 16500 & \\
	\bottomrule
    \end{tabularx}
    \caption{T-REX mountain wave test case: horizontal resolution \(\Delta x\), vertical resolution \(\Delta z\), and wall-clock times (WT) for the uniform mesh and the non-conforming meshes. The speed-up is computed considering the same maximum spatial resolution, i.e. comparing the wall-clock time of the uniform mesh and the wall-clock time of the coarsest non-conforming mesh (bold WT, see also main text for further details).}
    \label{tab:wall_clock_TREX}
\end{table}

\FloatBarrier

\subsection{3D medium-steep bell-shaped hill}
\label{ssec:3D_Melvin}

\begin{figure}[h!]
    \centering
    \begin{subfigure}{\textwidth}
	\centering
        \includegraphics[width = 0.7\textwidth]{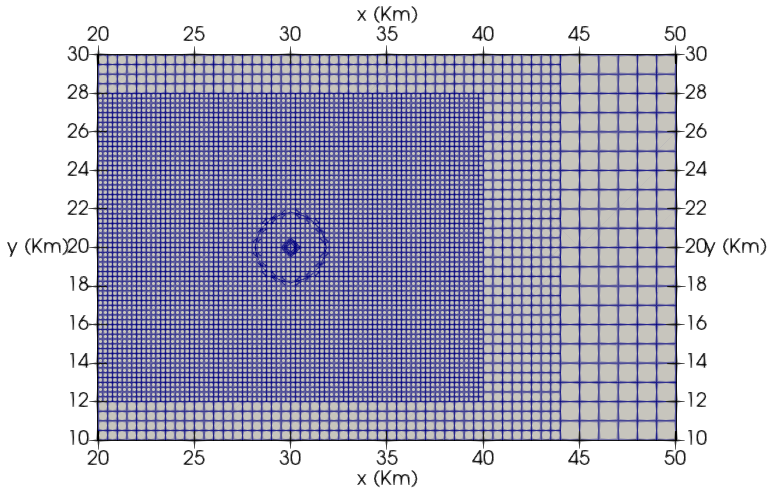}
    \end{subfigure}
    \begin{subfigure}{\textwidth}
	\centering
        \includegraphics[width = 0.7\textwidth]{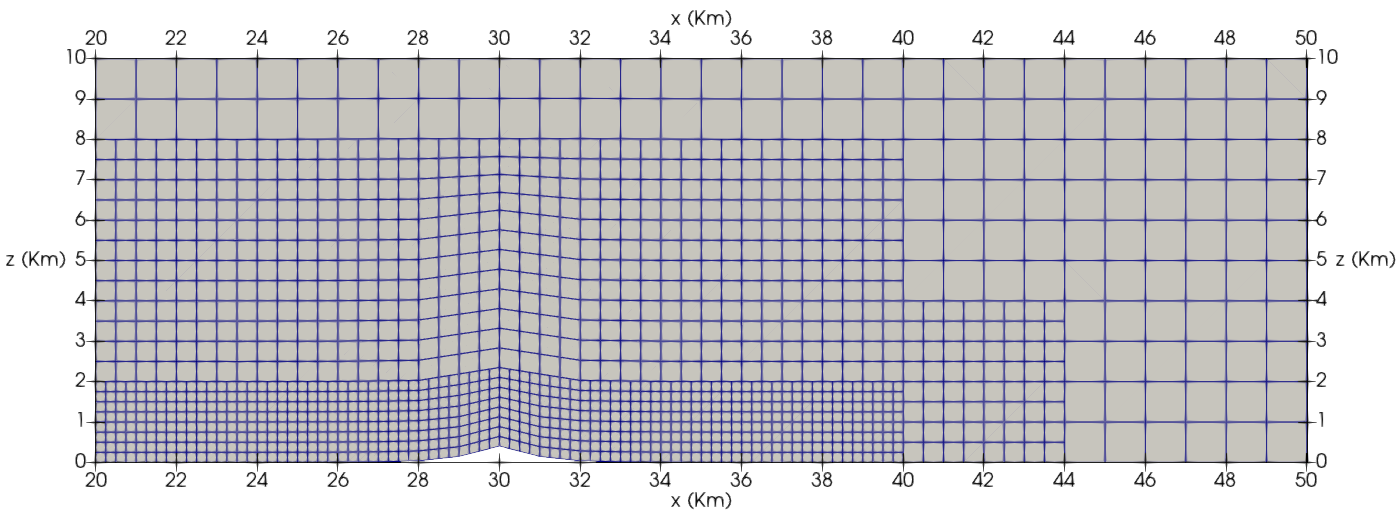}
    \end{subfigure}
    \caption{3D medium-steep bell-shaped hill test case, non conforming mesh. Left: \(x-y\) slice at \(z = \SI{800}{\meter}\). Right: \(x-z\) slice at \(y = \SI{20}{\kilo\meter}\).}
    \label{fig:3D_non_conforming_mesh}
\end{figure}

\begin{figure}[h!]
    \centering
    \begin{subfigure}{0.475\textwidth}
	\centering
        \includegraphics[width = 0.925\textwidth]  {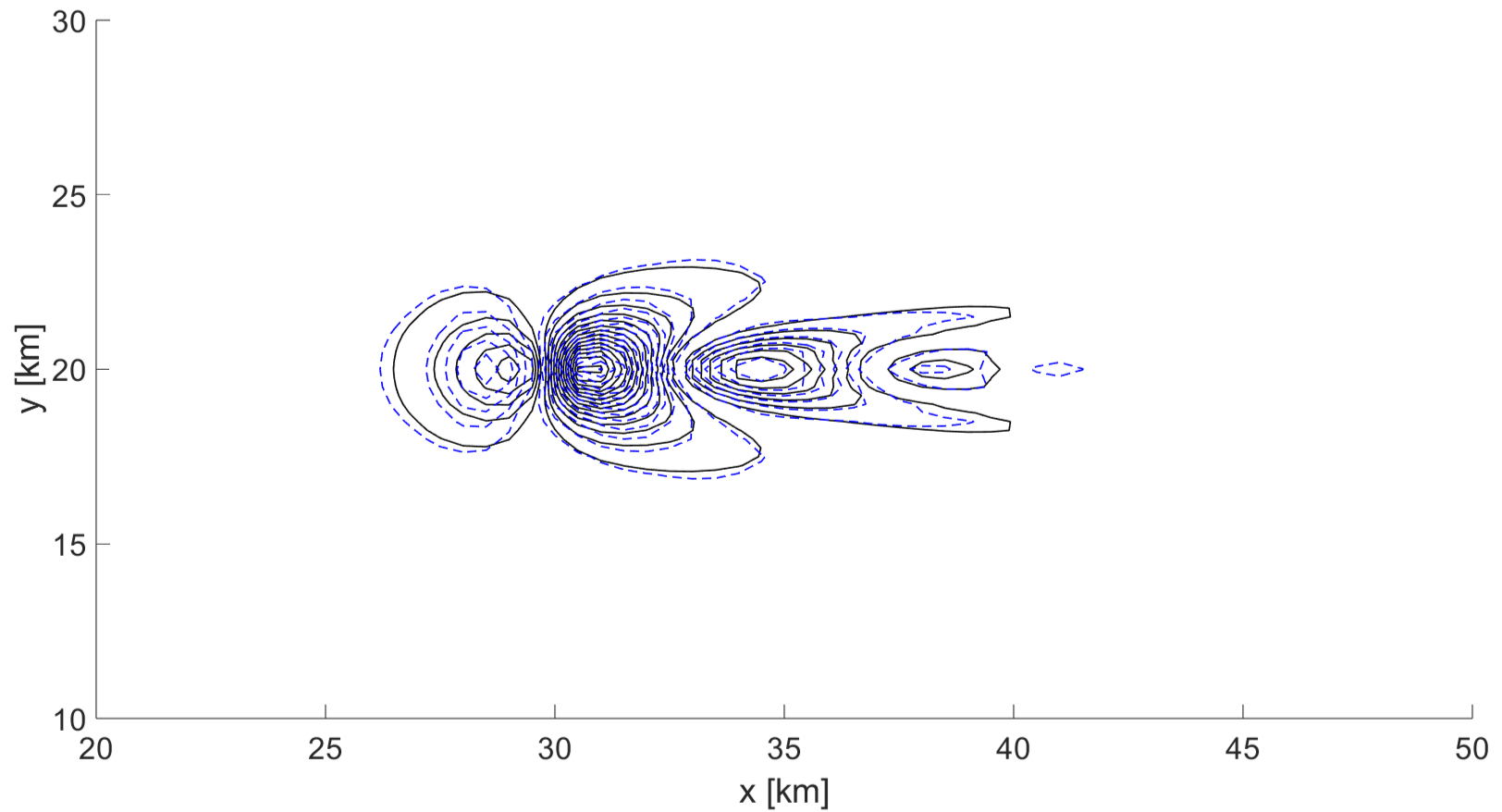} a)
    \end{subfigure}
    \begin{subfigure}{0.475\textwidth}
	\centering
        \includegraphics[width = 0.925\textwidth]{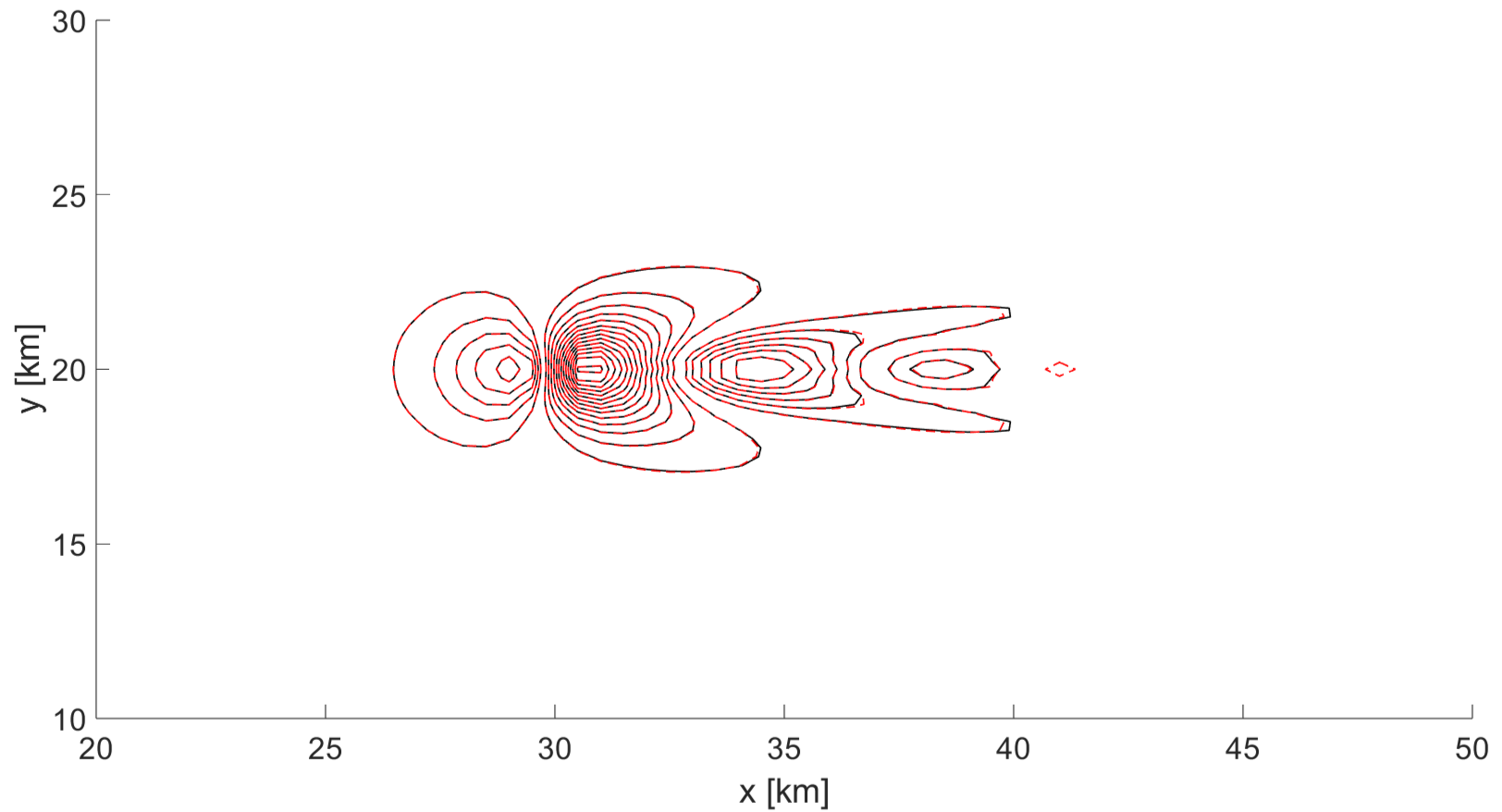} b)
    \end{subfigure}
    \begin{subfigure}{0.475\textwidth}
	\centering
        \includegraphics[width = 0.925\textwidth]{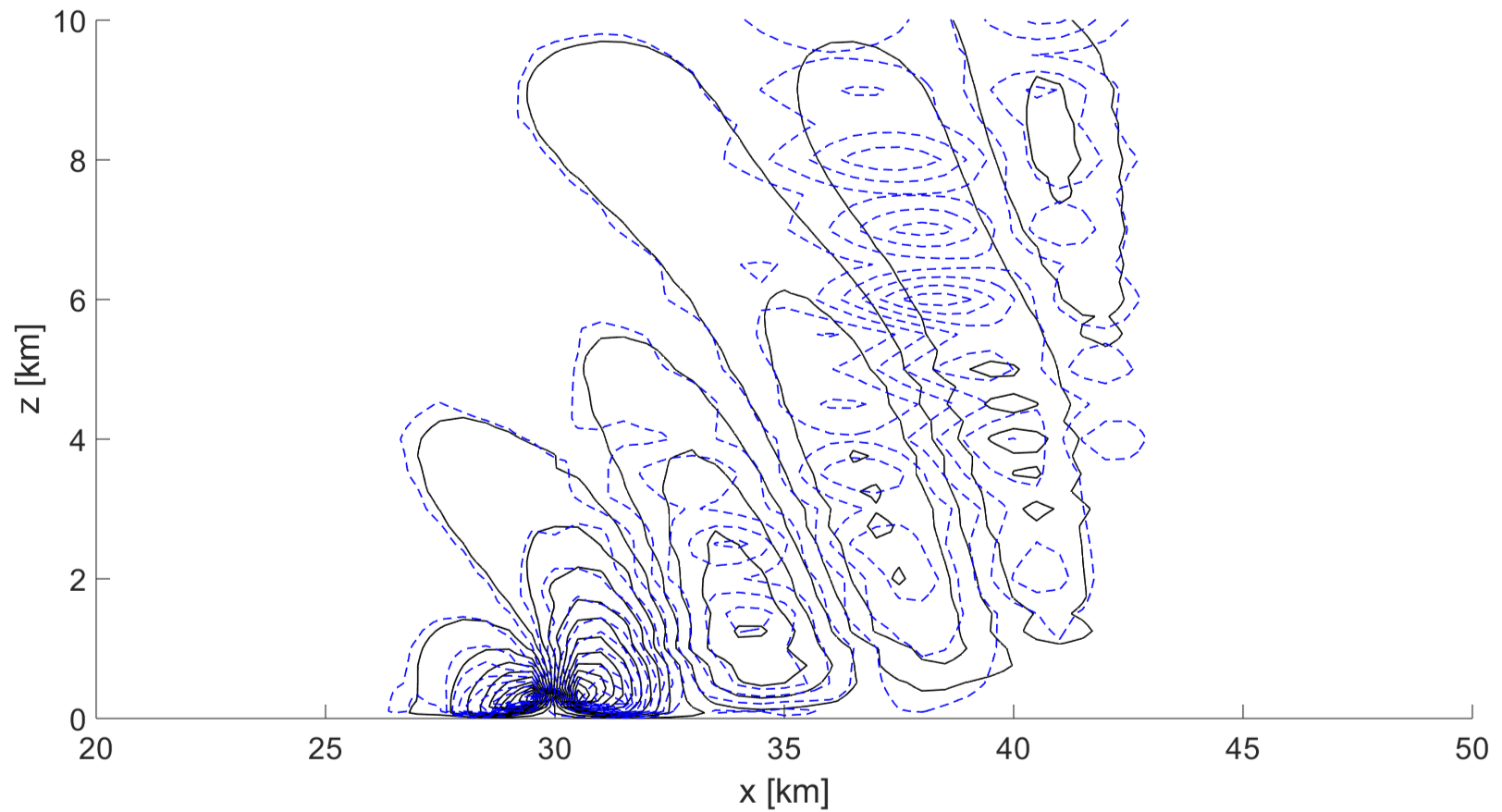} c)
    \end{subfigure}
    \begin{subfigure}{0.475\textwidth}
        \centering
        \includegraphics[width = 0.925\textwidth]{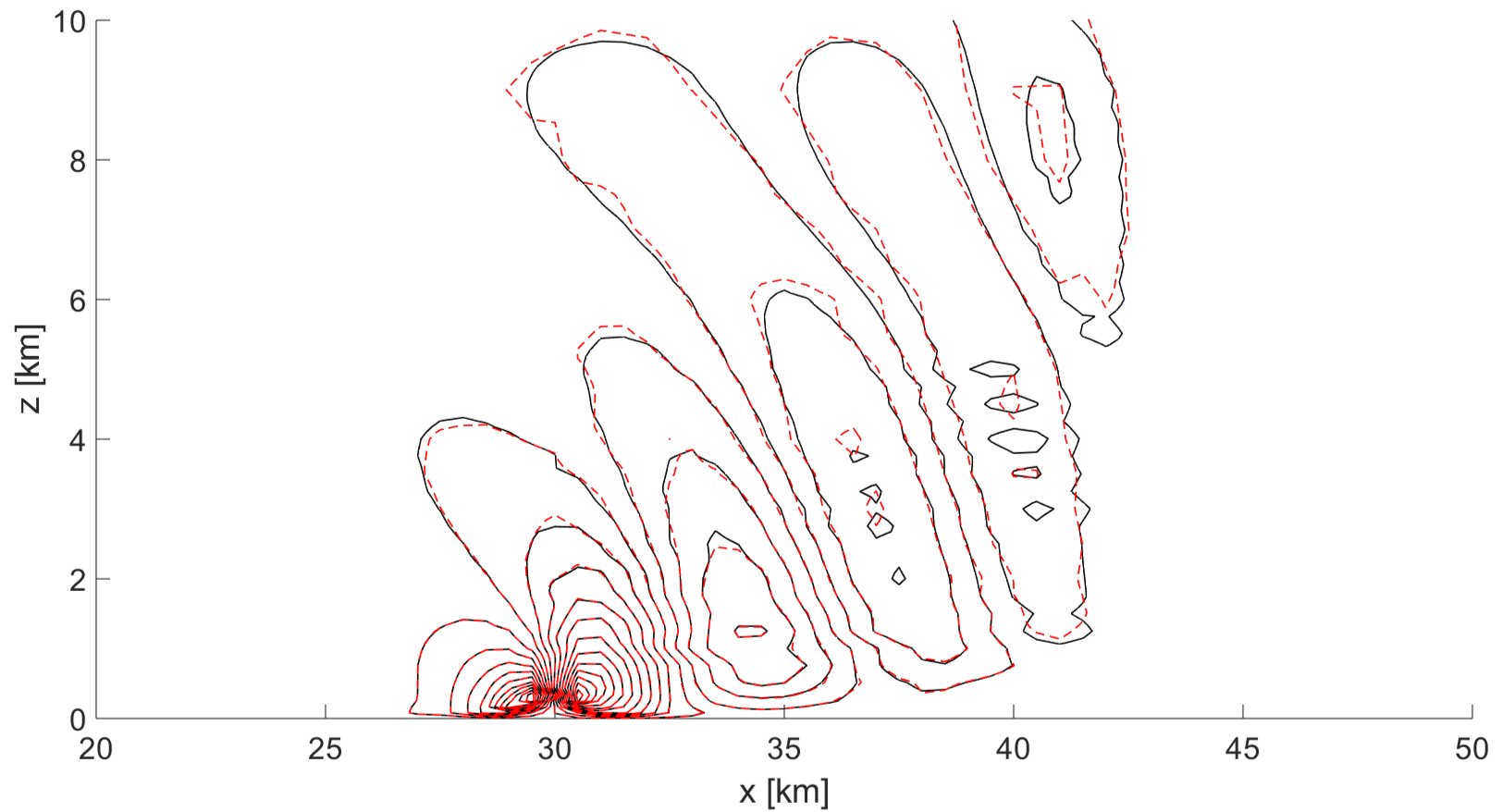} d)
    \end{subfigure}
    \caption{3D medium-steep bell-shaped hill test case at \(T_{f} = \SI{10}{\hour}\), vertical velocity contours. Top: \(x-y\) slice at \(z = \SI{800}{\meter}\) in the range\(\SI[parse-numbers=false]{[-1.5,1.3]}{\meter\per\second}\) with a \(\SI{0.1}{\meter\per\second}\) interval, a) comparison between the fine uniform mesh (black lines) and the coarse uniform mesh (blue lines), b) comparison between the fine uniform mesh (black lines) and the non-conforming mesh (red lines). Bottom: \(x-z\) slice at \(y = \SI{20}{\kilo\meter}\) in the range\(\SI[parse-numbers=false]{[-2.25,2]}{\meter\per\second}\) with a \(\SI{0.2}{\meter\per\second}\) interval, c) comparison between the fine uniform mesh (black lines) and the coarse uniform mesh (blue lines), d) comparison between the fine uniform mesh (black lines) and the non-conforming mesh (red lines).}
    \label{fig:3D_contours}
\end{figure}

Finally, we consider a three-dimensional configuration, focusing on the flow over a bell-shaped hill discussed e.g. in \cite{melvin:2019, orlando:2023a}, which we briefly recall here for the convenience of the reader. The computational domain is \(\Omega = \SI[parse-numbers=false]{\left(0, 60\right) \times \left(0, 40\right) \times \left(0, 16\right)}{\kilo\meter}\). The mountain profile is a 3D extension of the \textit{versiera of Agnesi} and can be defined as:
\begin{equation}
    h(x,y) = \frac{h_{c}}{\left[1 + \left(\frac{x - x_{c}}{a_{c}}\right)^{2} + \left(\frac{y - y_{c}}{a_{c}}\right)^{2}\right]^{\frac{3}{2}}},
\end{equation}
with \(h_{c} = \SI{400}{\meter}, a_{c} = \SI{1}{\kilo\meter}, x_{c} = \SI{30}{\kilo\meter},\) and \(y_{c} = \SI{20}{\kilo\meter}\). The buoyancy frequency is \(N = \SI{0.01}{\per\second}\), whereas the background velocity is \(\overline{u} = \SI{10}{\meter\per\second}\). Hence, since \(\frac{N a_{c}}{\overline{u}} = 1\), we are in a non-hydrostatic regime. The background potential temperature and Exner pressure profiles are those reported in Section \ref{ssec:nonhydrostatic} with \(\theta_{ref} = \SI{293.15}{\kelvin}\). The final time is \(T_{f} = \SI{1}{\hour}\). The damping layer is applied in the topmost \(\SI{6}{\kilo\meter}\) of the domain and in the first and last 20 km along the lateral boundaries with \(\overline{\lambda}\Delta t = 1.2\). We take polynomial degree \(r = 4\) and we consider three different computational meshes: a coarse uniform mesh composed by \(30 \times 20 \times 8 = 4800\) elements, i.e. a resolution of \(\SI{500}{\meter}\) along all the directions, a fine uniform mesh composed by \(60 \times 40 \times 16\) elements, i.e. a resolution of \(\SI{250}{\meter}\), and a non-conforming mesh with 3 different levels, composed by \(N_{el} = 1958\), with the finest level corresponding to the resolution of the finest uniform mesh (Figure \ref{fig:3D_non_conforming_mesh}). The time step is \(\Delta t = \SI{2}{\second}\), yielding a maximum acoustic Courant number \(C \approx 2.75\) and a maximum advective Courant number \(C_{u} \approx 0.13\) for the finest uniform mesh. The contours plots of the vertical velocity on a \(x-y\) slice placed at \(z= \SI{800}{\meter}\) and on a \(x-z\) slice placed at \(y = \SI{20}{\kilo\meter}\) show once more the accuracy and the robustness of simulations employing non-conforming meshes (Figure \ref{fig:3D_contours}). No spurious wave reflections arise at the internal boundaries that separate regions with different resolutions. Moreover, one can easily notice that the change of resolution affects the development of lee waves. However, it is sufficient to employ a higher resolution  only around the orography, whereas larger scales along all the directions can be resolved at a much coarser resolution. Thanks to its significantly lower number of degrees of freedom, the use of non-conforming mesh yields a computational time saving of around 15\% with respect to the coarse uniform mesh and of around 93\% with respect to the fine uniform mesh (Table \ref{tab:wall_clock_3D}).

\begin{table}[h!]
    \centering
    \footnotesize
    \begin{tabularx}{0.5\columnwidth}{rrrr}
	\toprule
        $N_{el}$ & $\Delta[\SI{}{\meter}]$ & WT$[\SI{}{\second}]$ & Speed-up \\
	\midrule
	4800 (uniform) & 500.0 & 3020 & \\
	\midrule
	38400 (uniform) & 250.0 & \textbf{36500} & \\
	\midrule
	1958 (non-conforming) & 250.0 & \textbf{2560} & \textbf{14} \\
	\bottomrule
    \end{tabularx}
    \caption{3D medium-steep bell-shaped hill test case: resolution \(\Delta\) and wall-clock times (WT) for the uniform meshes and the non-conforming mesh. The speed-up is computed considering the same maximum spatial resolution, i.e. comparing the wall-clock time of the finest uniform mesh and the wall-clock time of the non-conforming mesh (bold WT, see also main text for further details).}
    \label{tab:wall_clock_3D}
\end{table}

\section{Conclusions}
\label{sec:conclu} \indent

We have presented a systematic assessment of non-conforming meshes for the simulation of flows over orography using an IMEX-DG numerical model for the compressible Euler equations. For this purpose, we have exploited the adaptation framework provided by the open-source numerical library \texttt{deal.II} \cite{arndt:2023, bangerth:2007}. 
The proposed approach allows local mesh refinement both in the horizontal and vertical direction, without the need to apply relaxation procedures along the interfaces between the coarse and fine meshes. At a given accuracy level, the use of non-conforming meshes enables a significant reduction in the number of computational degrees of freedom with respect to uniform resolution meshes. The numerical results show that stable simulations are produced with no spurious reflections at internal boundaries separating mesh regions with different resolutions. In addition, accurate values for the momentum flux are retrieved in robust non-conforming simulations for increasingly realistic orography profiles. 

Numerical simulations with non-conforming meshes can use substantially higher resolution only near the orographic features, correctly reproducing the larger scale, far-field orographic response, while using meshes that are relatively coarse over most of the domain. 
In a context of spatial resolutions approaching the hectometric scale in numerical weather prediction models, these results support the use of locally refined, non-conforming meshes as a reliable and effective tool to greatly reduce the dependence of atmospheric models on orographic wave drag parametrizations. Indeed, the results obtained in our framework envisage the use of locally refined, non-conforming meshes as a reliable, effective tool to push NWP and climate models out of the `grey zone' with respect to the resolution of orographic effects \cite{kanehama:2019, sandu:2019}.

In future developments, we will implement specific multilevel preconditioners in the matrix-free approach of the \texttt{deal.II} library, in order to get the full benefit from the significant reduction in number of degrees of freedom allowed by the use of non-conforming meshes for more realistic configurations. We also plan to consider the inclusion of more complex physical phenomena, such as more sophisticated turbulence models, water vapour transport, and adiabatic heating, as well as exploring physics-dynamics coupling, in order to demonstrate that all the typical features of a high resolution numerical weather prediction model can be included in the proposed adaptive framework without loss of accuracy. Moreover, the proper thermodynamic description of atmosphere dynamics is becoming a matter of deep investigation \cite{staniforth:2022}. The assumption of ideal gas for dry air and water vapour \cite{staniforth:2019} is not always a proper one, especially if phase changes occur. Recent work by two of the authors \cite{orlando:2022} can handle more general equations of state for real gases, thus paving the way to the inclusion of effects due to water vapour and moist species in a more realistic framework.

\section*{Acknowledgements}

The simulations have been partly run at CINECA thanks to the computational resources made available through the ISCRA-C projects FEMTUF - HP10CTQ8X7 and FEM-GPU - HP10CQYKJ1 and through the EuroHPC JU Benchmark And Development project EHPC-BEN-2024B03-045. This work has been partly supported by the ESCAPE-2 project, European Union’s Horizon 2020 Research and Innovation Programme (Grant Agreement No. 800897). We would also like to thank Dr. Christian K{\"u}hnlein, Dr. James Doyle and Dr. Sa\u sa Gaber\u sek for providing the data for the T-REX mountain wave.

\printbibliography

\end{document}